\documentclass[useAMS,usenatbib,longabstract]{mn2e}
\usepackage{graphicx}
\usepackage{amssymb}
\begin{document}

\title[Low Luminosity Type II Supernovae]
{
Underluminous Type II Plateau Supernovae: \\
II. Pointing towards moderate mass precursors
  }

\author[S. Spiro et al.]
        {S. Spiro$^{1}$, A. Pastorello$^{1}$, M.L. Pumo$^{1,2}$, L. Zampieri$^{1}$, M. Turatto$^{1}$,  S.J. Smartt$^{3}$, \and
        S. Benetti$^{1}$, E. Cappellaro$^{1}$,  S. Valenti$^{4,5}$,   I. Agnoletto $^{1}$, G. Altavilla$^{6}$, \and
        T. Aoki $^{7}$, E. Brocato $^{8}$, E.M. Corsini$^{9,1}$, 
        A. Di Cianno$^{10}$, N. Elias-Rosa$^{11}$, \and
        M. Hamuy$^{14}$, K. Enya $^{12}$, M. Fiaschi$^{9}$, G. Folatelli$^{13}$, S. Desidera$^{1}$, A. Harutyunyan$^{15}$, \and
         D. A. Howell$^{4,5}$,  A. Kawka$^{16}$, Y. Kobayashi $^{17}$, B. Leibundgut$^{18}$,\and
           T. Minezaki $^{7}$,  H. Navasardyan$^{1}$,  K. Nomoto $^{19, 20}$, S.Mattila$^{21}$, A. Pietrinferni$^{10}$,\and
          G. Pignata$^{22}$,  G. Raimondo$^{10}$,  M. Salvo$^{23}$, B.P. Schmidt$^{23}$, J. Sollerman$^{24}$, \and
           J. Spyromilio$^{18}$, S. Taubenberger$^{25}$, G. Valentini $^{10}$, S. Vennes$^{16}$, 
          Y. Yoshii $^{7}$ \\
          $^{1}$ INAF - Astronomical Observatory of Padua, Vicolo dell'Osservatorio 5, 35122 Padua, Italy \\
          $^{2}$ INAF - Astrophysical Observatory of Catania, via S. Sofia 78, 95123 Catania, Italy \\
          $^{3}$ Astrophysics Research Centre, School of mathematics and Physics, Queen's University Belfast, Belfast BT7 INN, United Kingdom\\ 
	 $^{4}$ Las Cumbres Observatory Global Telescope Network, 6740 Cortona Dr., Suite 102, Goleta, CA 93117, USA\\
          $^{5}$ Department of Physics, University of California, Santa Barbara, Broida Hall, Mail Code 9530, Santa Barbara, CA 93106-9530, USA\\
	 $^{6}$ Osservatorio Astronomico di Bologna, INAF, Via C. Ranzani, 1, I-40127 Bologna, Italy \\
          $^{7}$ Kiso Observatory, Institute of Astronomy, School of Science, The University of Tokyo, 10762-30 Mitake, Kiso, Nagano 397-0101, Japan\\
	 $^{8}$ INAF-Osservatorio Astronomico di Roma, Via Frascati 33, 00040, Monte Porzio Catone, Italy\\
	$^{9}$ Dipartimento di Fisica e Astronomia `G. Galilei', Universit\'a di Padova, vicolo dell'Osservatorio 3, I-35122 Padova, Italy \\
	 $^{10}$ INAF-Osservatorio Astronomico di Teramo, via M. Maggini, I-64100 Teramo, Italy\\
          $^{11}$ Institut de Ci\`encies de l'Espai (IEEC-CSIC), Facultat de Ci\`encies, Campus UAB, 08193 Bellaterra, Spain\\
          $^{12}$ Institute of Space and Astronautical Science, Japan Aerospace Exploration Agency, 3-1-1, Yoshinodai, Sagamihara, Kanagawa 229-8510, Japan \\
	 $^{13}$ Kavli Institute for the Physics and Mathematics of the Universe (WPI), Todai Institutes for Advanced Study, the University of Tokyo,\\
	  Kashiwa, Japan 277-8583 (Kavli IPMU, WPI)\\
	 $^{14}$ Departamento de Astronom'a, Universidad de Chile, Casilla 36-D, Santiago, Chile \\
	 $^{15}$ Fundacion Galileo Galilei - INAF, Telescopio Nazionale Galileo, 38700 Santa Cruz de la Palma, Tenerife, Spain\\
	 $^{16}$ Astronomick\'y \'ustav, AV C$\check{R}$, Fri$\check{c}$ova 298, CZ-251 65 Ond$\check{r}$ejov, Czech Republic \\
	 $^{17}$ National Astronomical Observatory, 2-21-1 Osawa, Miaka, Tokyo 181-8588, Japan\\
	 $^{18}$ ESO, Karl-Schwarzschild-Strasse 2, D-85748 Garching, Germany \\ 
	 $^{19}$ Kavli Institute for the Physics and Mathematics of the Universe (WPI), The University of Tokyo, 5-1-5 Kashiwanoha, Kashiwa,\\
	  Chiba 277-8583, Japan\\
	 $^{20}$ Department of Astronomy, School of Science, The University of Tokyo, 7-3-1 Hongo, Bunkyo-ku, Tokyo 113-0033, Japan\\
	 $^{21}$ Finnish Centre for Astronomy with ESO (FINCA), University of Turku, V\"ais\"al\"antie 20, FI-21500 Piikki\"o, Finland \\
	 $^{22}$ Departamento de Ciencias Fisicas, Universidad Andres Bello, Avda. Republica 252, Santiago, Chile \\
	 $^{23}$ Research School of Astronomy and Astrophysics, Australian National University, Cotter Road, Weston Creek, PO 2611, Australia \\
	 $^{24}$ Oskar Klein Centre, Department of Astronomy, AlbaNova, Stockholm University, SE-10691, Stockholm, Sweden\\
	 $^{25}$ Max-Planck-Institut f\"ur Astrophysik, Karl-Schwarzschild-Str. 1, D-85741, Garching bei M\"unchen, Germany\\	 
        }
\date{Accepted .....;
      Received ....;
      in original form ....}
\maketitle
\clearpage
\begin{abstract}
  We present new data for five under-luminous type
  II-plateau supernovae (SNe IIP), namely SN 1999gn, SN 2002gd, SN
  2003Z, SN 2004eg and SN 2006ov. This new sample of low-luminosity
  SNe IIP (LL SNe IIP) is analyzed together with similar objects
  studied in the past.
  All of them show a flat light curve
  plateau lasting about 100 days, an under luminous late-time
  exponential tail, intrinsic colours that are unusually red,
  and spectra showing prominent and narrow P--Cygni lines. 
A velocity of the ejected material below 10$^3$ km s$^{-1}$ is
inferred from measurements at the end of the plateau.  The $^{56}$Ni  masses ejected in the explosion are very small ($\leq$ 10$^{-2}$ M$_\odot$).
  We investigate the correlations among $^{56}$Ni mass, expansion
  velocity of the ejecta and absolute magnitude in the middle of the
  plateau, confirming the main findings of \cite{Hamuy03}, according
  to which events showing brighter plateau and larger expansion
  velocities are expected to produce more $^{56}$Ni.  We propose that
  these faint objects represent the low luminosity tail of a
  continuous distribution in parameters space of SNe IIP.  The
  physical properties of the progenitors at the explosion are
  estimated through the hydrodynamical modeling of the observables for
  two representative events of this class, namely SN 2005cs and SN
  2008in.  
  We find that the majority of LL SNe
  IIP, and quite possibly all,  originate in the core-collapse of intermediate mass
  stars, in the mass range 10-15 M$_{\odot}$. 
\end{abstract}


\section{Introduction}

In recent years, 
a number of under-luminous transient events observed in nearby galaxies, with 
 peculiar observed properties have provoked interest in their physical
 origins and explosion mechanisms. 
They lie in the gap of the absolute 
magnitude vs. photometric evolutionary timescales diagram ($-10 >M_R> -14$ mag, see \citealp{Kulkarni07}; 
\citealp{Rau09}) separating the brightest novae and the faintest known supernovae (SNe).
These new discoveries include a number of faint, $^{56}$Ni-poor, low expansion velocity
hydrogen free  SNe (e.g. SN 2008ha, \citealp{Valenti09}; 
\citealp{Foley09}), some ultra-faint objects with type IIn SN-like
spectra and shape of the light curves resembling those of type IIL or type IIP SNe
(NGC300 OT2008-1, \citealp{Bond09}; \citealp{Berger09}; 
SN 2008S, \citealp{Botticella09}; \citealp{Smith09}; M85 OT2006-1, 
\citealp{Kulkarni07}; \citealp{Pastorello07}), and
even a sub-luminous SN 1987A-like event (SN 2009E, \citealp{Pastorello12}).

Whilst the nature of many of these objects is still far from being unveiled (unusual eruptive events or real SNe;
partial deflagrations or low-energy genuine SNe; electron-capture SNe
from moderate-mass stars or fall-back core-collapse supernovae from massive progenitors), 
major progresses has been made during the past decade 
for a specific class of under-luminous transients: the faint type IIP SNe
(\citealp{Pastorello04}).

The prototype  of this class is SN 1997D. 
It was discovered on January 14.15 UT in NGC 1536 (\citealp{DeMello97})
 and it was at that time the faintest and the most sub energetic Type II SN
ever discovered. The comparison of the light curve with those of other Type II SNe, suggested the ejection of very low amount 
of $^{56}$Ni (\citealp{Turatto98}, \citealp{Benetti01}).
Unfortunately SN 1997D was discovered when it was quite old, so the explosion epoch was not well constrained. It had unusually red spectra suggesting 
low continuum temperatures, and narrow spectral
lines indicative of slow-moving ejecta (1000 km s$^{-1}$, that is 3-4 times
lower than in normal type IIP SN explosions). All of this suggested
a rather low explosion energy for SN 1997D (of the order of 10$^{-1}$ foe).

 In past years, two alternative scenarios have been proposed to explain SN 1997D. 
The first invokes a low-energy explosion of a high mass (more than 25 M$_\odot$) star 
(\citealp{Turatto98}) in which a large
amount of stellar material remains bound to the core after the collapse, 
and falls back onto it, increasing its
mass (\citealp{Zampieri98}). In this case the compact remnant may be either
a black hole or a neutron star. The crucial parameter to discriminate between 
the two types of remnants is indeed the amount of stellar
material falling-back onto the core.
An alternative scenario was proposed by \cite{Chugai00}, in which the 
precursor star of moderate mass ($\sim$8-11 M$_{\odot}$) develops a degenerate NeO core (see e.g.  \citealp{Pumo07}, \citealp{Pumo09} and references therein, for a discussion on the evolution of this stellar type) and undergoes a core collapse triggered by electron capture (\citealp{Nomoto84,Takahashi13}). In this case the fall back is negligible and the explosion leaves behind a neutron star. \\

As mentioned above, the explosion epoch of SN 1997D was poorly constrained.
Several models assumed a plateau duration of about 40-60 days
(e.g. \citealp{Chugai00}), while others a more conventional
length of about 100 days (e.g. \citealp{Zampieri03}).

Subsequent discoveries of a number of SN 1997D-like events established 
 that most of them had long-lasting plateau ($\sim$ 100 days), implying
that the results obtained from the preliminary modelling to the data of SN 1997D
needed to be revised.
 Several objects were indeed discovered very close to the explosion epoch 
and proved that their plateaus were significantly longer than thought before
(e.g SNe 1999br, 2001dc and 2003Z, \citealp{Pastorello04}). 
These new observations allowed a consistent calibration of the explosion dates
also for events discovered at later phases, such as SN 1997D,  
and hence their explosion and ejecta parameters were better constrained.

With more complete data sets and/or recalibrated explosion epochs, 
the masses inferred from hydrodynamic models were initially estimated to be relatively high. 
\cite{Zampieri03} found a progenitor of about 16 M$_{\odot}$ 
 for SN 1999br  and 19 M$_{\odot}$ for SN 1997D.  Similar results were obtained by 
\cite{Utrobin07} with a mass in the range 14.4-17.4  M$_{\odot}$ for the progenitor 
of SN 2003Z, 
while \cite{Utrobin08} and \cite{Pastorello09} 
estimated the red supergiant precursor of SN 2005cs to be 
18.2$\pm$1.0 M$_{\odot}$  and 10-15 M$_{\odot}$, respectively. Therefore, based
on SN data modelling, the masses of low-luminosity SNe IIP progenitor
were estimated in the range 10-19 M$_{\odot}$.

An independent estimate of the masses of the SN progenitor is based on 
the direct detection of the precursor star in archive images obtained before the SN explosion 
(\citealp{VanDyk03, Smartt04, Smartt09, Smartt09b} and reference therein).   
The detection of a progenitor (or the measurement of deep upper magnitude limits) on images obtained 
with different filters allows us to estimate the 
absolute luminosity and the colours (and, from these, the temperature) of the putative 
progenitor star. Hence, from matching theoretical evolutionary tracks, it is possible to estimate its mass.
In the last decade a handful of LL SNe IIP have been studied with this method: SN 2005cs (\citealp{Maund05, Li06, Eldridge07, Maund13b}), 
SN 2008bk (Pignata et al. in prep., \citealp{Mattila08, VanDyk12, Maund13}) and SN 2009md (\citealp{Fraser11}) for which lower mass progenitors 
of 8-13 M$_{\odot}$ were inferred.
For two additional LL SNe IIP, SN 1999br (\citealp{Maund05b, Smartt09}) and SN 2006ov (\citealp{Crockett11}), progenitor mass limits of M $<$ 15 M$_{\odot}$ 
and M $<$ 10 M$_{\odot}$, respectively, were estimated.

Currently the masses of precursors of LL SNe IIP are
still uncertain, with conflicting results obtained through 
hydrodynamic modelling of SN data,
which indicates a distribution extending to rather massive progenitors ($\sim$10-20 M$_\odot$)
or from the direct detection of progenitors in pre-SN images, which suggests lower 
mass ($\sim$8-13 M$_\odot$) stars. 
The problem of this discrepancy has been raised by \cite{Utrobin09}, that  
tentatively attributed an over-estimate
of the ejecta masses to the one-dimensional approximation of the
hydrodynamical models.  \cite{Dessart13}  suggested it may be related
to the pre-supernova structure being not well understood.

On the other hand it is fair to say that estimate from the 
direct detection of the progenitor presents a number of caveats. 
First of all this method can be applied only to nearby objects within a small volume, 
e.g. those that explodes within $\sim$30 Mpc. 
Uncertainties in the stellar evolutionary models used to
infer the progenitor mass (e.g. treatment of
overshooting and mass loss) can also lead to different inferred
masses,  although none of the codes available differ by large factors
\citealp{Smartt09}.
Finally reliable extinction estimates are essential and there is a 
risk of underestimating the mass of the progenitor
(\citealp{Walmswell12}). 
\citet{Kochanek2012} also note that the extinction by circumstellar dust may be significantly different from that of interstellar dust and that 
 the effect of 
CSM dust may not raise progenitor estimates as significantly as 
proposed by \citet{Walmswell12}.
Systematic studies of a larger number of well-followed faint SNe IIP will allow to
improve our knowledge of these events and, in particular,  of the parameters describing the progenitor star and its explosion
(including $^{56}$Ni  mass, total ejected mass, initial radius, kinetic energy).  \\

In this paper we present new data (both photometry and spectroscopy)
for five sub-luminous supernovae classified as LL SNe IIP: SN 2002gd, SN 2003Z, SN
2004eg, SN 1999gn and SN 2006ov. These supernovae, in analogy with SN 1997D, are intrinsically
faint compared with more canonical SNe IIP (the typical luminosity in the plateau phase for a normal type IIP is L $\sim$ 10$^{42}$ - 10$^{43}$ erg s$^{-1}$).  
Together  with other LL SNe IIP presented in 
previous works i.e. SN 1994N, SN 1997D, SN 1999br, SN 1999eu, SN 2001dc, SN 2005cs 
(\citealp{Turatto98, Benetti01, Pastorello04}, 2006, 2009) and other events 
recently published, i.e. SN 2008bk (Pignata et al. in prep., \citealp{VanDyk12}), SN 2008in (\citealp{Roy11}), 
SN 2009N (\citealp{Takats13}), SN 2009md (\citealp{Fraser11}) and SN 2010id (\citealp{GalYam11}),  we collect
an extensive sample of LL SNe IIP. 
As  the best observed 
underluminous SN IIP, SN 2005cs (\citealp{Pastorello06}, 2009; \citealp{Tsvetkov06, Brown07, 
Dessart08})  can be considered as a template for this SN family. 
In particular, the multi-wavelength study of SN 2005cs is useful to
derive the bolometric light curves of all other LL SNe IIP (with
caveats described in the text). 
In addition, the properties of its red supergiant progenitor star were extensively
discussed (\citealp{Maund05, Li06, Takats06, 
Eldridge07, Dessart08, Utrobin08}), and -likely- well
constrained. These results will be compared with our findings for other LL SNe IIP.\\

This paper is organized as follows: 
in section 2  we present the new LL SN IIP sample and introduce the properties of 
the host galaxies. In section 3 we give a summary of the observations 
 with a brief description of the instruments used. Photometric data are  presented 
in section 4, including light curves  (section 4.1), 
colour curves (section 4.2) and bolometric light curves (section
4.3). Spectroscopic data
are shown 
in section 5: properties of individual objects
are illustrated in section 5.1, while the common 
properties for the entire group are discussed in section 5.2.  
Estimates of the physical parameters of our SN sample and a discussion on the implications for the nature 
of the progenitor stars are in Sections 6 and 7, respectively.
Finally a summary follows in section 8.

\section{The new sample: supernovae and host galaxies}

We present here relevant data for the new LL SNe sample and their host galaxies. 
Except in the case of SN1999gn and SN 2006ov (see section 2.4) for which a Tully-Fisher distance is adopted, 
in all other cases the supernova distances are	 estimated using the recessional velocity corrected for Local Group infall into the Virgo Cluster (v$_{Vir}$ parameter
from the HYPERLEDA database\footnote{\it http://leda.univ-lyon1.fr/}) assuming an Hubble Constant H$_0$ = 72 km s$^{-1}$ Mpc$^{-1}$. 

No evidence of extinction within the host galaxies were found, therefore we consider only the Galactic contribution estimated by  \cite{Schlafly11}.

\subsection{SN 2002gd}

\begin{figure}
\begin{center}{
\includegraphics[width=8.5cm]{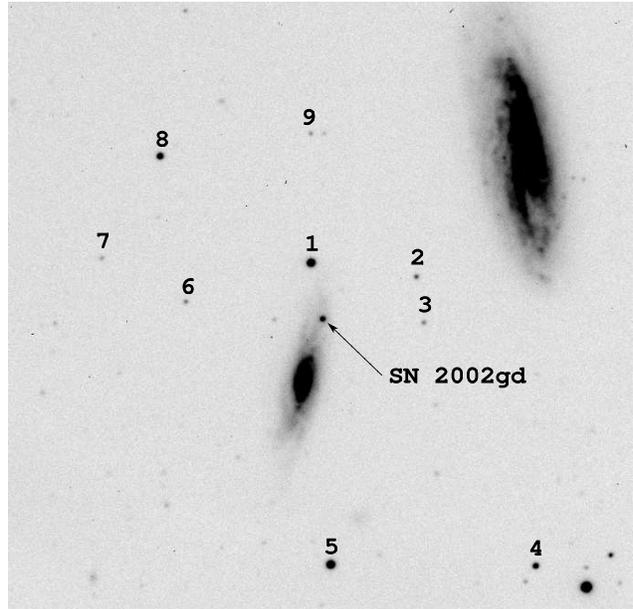}
\caption[SN 2002gd in NGC 7537]{SN 2002gd and the local sequence stars (TNG + Dolores, R band image 
obtained on 2002 December 28, with an exposure time of 3 minutes). East is up, North is to the right.} \label{sn02gd_field}}
\end{center}
\end{figure}

SN 2002gd (Fig. \ref{sn02gd_field}) was independently discovered by Klotz, Puckett \& Langoussis (2002). 
The earliest detection was recorded on 2002 October 5.31 UT at an unfiltered CCD magnitude of 18.9 (\citealp{Klotz02}).
Nothing was visible at the position of the SN on September 15.95 UT (\citealp{Klotz02}), indicating that
the SN was caught very young, during the steep rise to the maximum light.
Puckett provided the position of the SN at $\alpha$ = 23$^h$14$^m$36\fs98 
and $\delta$ = +04$\degr$30$\arcmin$05$\farcs$7
(equinox 2000.0), that was 36$\farcs$8 East and 10$\farcs$8 North of the nucleus of the galaxy NGC 7537. 
The SN was spectroscopically classified as a young SN II with a very blue continuum and H Balmer and He I 
5876 {\r{A}}  lines showing low--contrast P--Cygni profiles. The position of the blue-shifted minima  
 provide an expansion velocity v $\sim$ 5000 km s$^{-1}$ (\citealp{Hamuy02}), which is relatively low for such an early 
epoch (\citealp{Benetti02, Filippenko02}).
From the characteristics of the early time spectra and light curve
evolution, we can constrain the explosion epoch with a small
uncertainty to be 
2002 October 4th, i.e. JD = 2452552 $\pm$ 2. 

The host galaxy of SN 2002gd, NGC 7537, is of Sbc type.
Relevant data for SN 2002gd and the host galaxy are reported in Tab.\ref{datagal_gd}.

\subsection{SN 2003Z}

\begin{figure}
\begin{center}
\includegraphics[width=8.5cm]{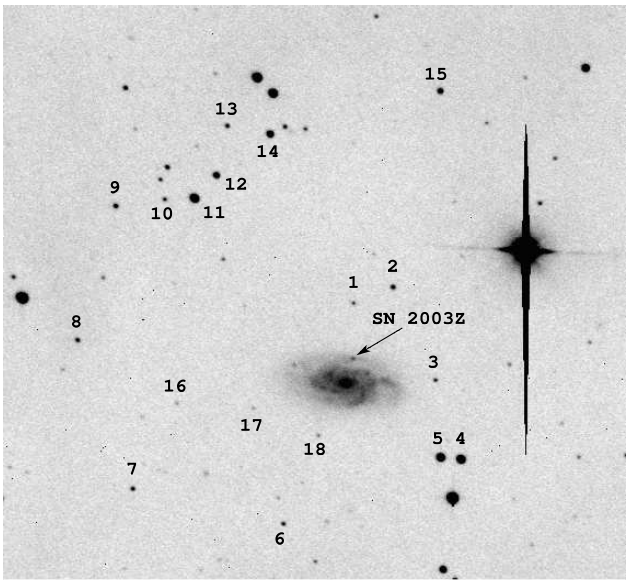}
\caption[SN 2003Z in NGC 2742]{SN 2003Z in NGC 2742 and the local sequence stars (R band image obtained on 2003 May 21 
with the Newton 0.4m telescope
of the {\it Gruppo Astrofili di Padova} (Italy)). North is up, East is to the left.} \label{sn03Z_field}
\end{center}
\end{figure}

SN~2003Z (Fig. \ref{sn03Z_field}) was discovered by Qiu $\&$ Hu (\citealp{Boles03}) with the Beijing Astronomical Observatory (BAO) 0.6m telescope on
2003 January 29.7, when the SN magnitude was $\sim$ 16.7, and  confirmed with Katzman Automatic Imaging Telescope (KAIT) on 2003 January 30.4 (magnitude 16.5).
It was located at $\alpha$ = 09$^h$07$^m$32\fs46  and $\delta$ = +60$\degr$29$\arcmin$17$\farcs$5 (equinox 2000.0),
close to an arm of NGC 2742 (8$^{\prime\prime}$.4 West and 31$^{\prime\prime}$.0 North
of the nucleus of the host galaxy). Another BAO image obtained on 2003 January 20.7 showed nothing at the SN position,
suggesting that SN~2003Z was discovered shortly after the explosion (JD = 2452665 $\pm$ 4.5). 
\citealp{Matheson03} obtained a spectrum on 2003 January 31.36 showing a very blue 
continuum with well--defined P--Cygni lines of H and HeI. The estimated expansion velocity as derived
from the minimum of H${\beta}$ was 6800 km s$^{-1}$ (\citealp{Matheson03}), again rather low in comparison with the velocity of canonical 
SNe IIP a few days after the explosion.

NGC 2742 belongs to the small group of galaxies LGG 167 (\citealp{Garcia93}) that has a mean radial velocity of
1475 km s$^{-1}$. The peculiar velocity of NGC 2742 inside the group is +66 km s$^{-1}$. The correction for peculiar motion was taken into account in the distance modulus estimation.

Information on SN 2003Z and its host galaxy is in Tab.\ref{datagal_gd}.

\subsection{SN 2004eg}

\begin{figure}
\begin{center}{
\includegraphics[width=8.5cm]{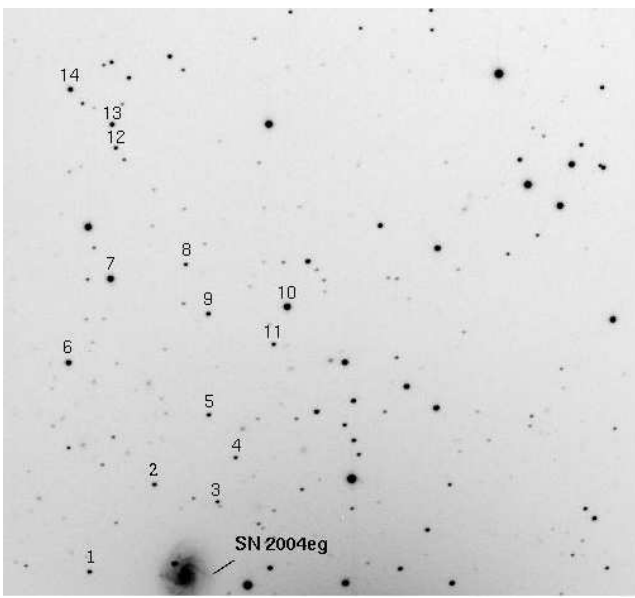}
\caption[SN 2004eg in UGC 3053]{Position of SN 2004eg in UGC 3053 and 
 local sequence stars (V band image obtained on 2004 December 1st 
with the 1.82-m Copernico telescope of Mt. Ekar, Asiago (Italy)). North is up, East is to the left.
} \label{sn04eg_field}}
\end{center}
\end{figure}

SN 2004eg (Fig. \ref{sn04eg_field}) was first detected in UGC 3053 by Young using the Table Mountain observatory 0.60-m reflector on 2004 September 1.488 UT  (\citealp{Young04}). The SN was located at $\alpha
= 04^{h} 28^{m} 08^{s}.26$, $\delta$ = +21$\degr$ 39$\arcmin$ 18$\farcs$3  (equinox
2000.0), 20$^{\prime\prime}$ West  and 1$^{\prime\prime}$.4 South of the nucleus of UGC
3053. The SN was not present on the Sky Survey images dating 1988-1991
(limiting  red mag 20.8-21.5; limiting blue mag 21.5-22.5) and it was 
classified as a type II SN by \cite{Filippenko04}.
The line velocity in the classification spectrum was found to be extremely low, 
roughly 500 km s$^{-1}$ (\citealp{Filippenko04}), as
measured from the position of the absorption minima of the P--Cygni profiles.

The adopted explosion epoch  is JD = 2453170 $\pm$ 30 inferred from the colour evolution (see section 4.2).
Main data of SN 2004eg and its host galaxy UGC 3053 are in Tab.\ref{datagal_gd}.

\subsection{SN 1999gn and SN 2006ov in M61}

M61 (NGC 4303) is one of the largest galaxies in the Virgo cluster; its 6' diameter correspond 
to about 100,000 light years, similar to the diameter of the Milky Way. To date,
6 SNe have been registered in this galaxy : SN 1926A (type IIL, mag 12.8), 
SN 1961I (type II, mag 13), SN 1964F (type I, mag 12), SN 1999gn and SN 2006ov 
(type IIP, this paper) and, more recently, the type IIP SN 2008in (see Sect. 7).
A Tully-Fisher distance d = 12.6 $\pm$ 2.4 Mpc ($\mu$ = 30.5 $\pm$ 0.4) 
is adopted for M61, in agreement with \cite{Li07} and  \cite{Smartt09}.

SN 1999gn was discovered with a
0.50-m telescope by A. Dimai on 1999 december 17.22 UT at $\alpha
= 12^{h} 21^{m} 57^{s}.02$, $\delta$ = +04$\degr$ 27$\arcmin$ 45$\farcs6$
(31$^{\prime\prime}$.7 East and 39$^{\prime\prime}$.8 South of the nucleus of the galaxy),  when the object 
had magnitude 16.0. The new object was confirmed on the subsequent day through a KAIT image 
at a magnitude of about 15.5. 
The supernova was clearly discovered in the rising phase to the maximum light 
(\citealp{Dimai99}). In a KAIT image taken on Nov. 26.5
no source was detected at the position of the transient to a limiting mag of about 19.0.
The spectroscopic classification of SN 1999gn as type II 
was assigned by \cite{Ayani99} from a wide-band spectrum (range 400-800 nm, 
resolution 0.6 nm) obtained with the Bisei Astronomical Observatory 1.01-m telescope. 
The spectrum exhibited a blue continuum with overimposed broad H$\alpha$ and H$\beta$ 
P--Cygni profiles. 
The expansion velocity of the supernova was 5300 km/s, as derived from the position of the 
H$\alpha$ absorption minimum (\citealp{Ayani99}).

\begin{figure}
\begin{center}{
\includegraphics[width=8.5cm]{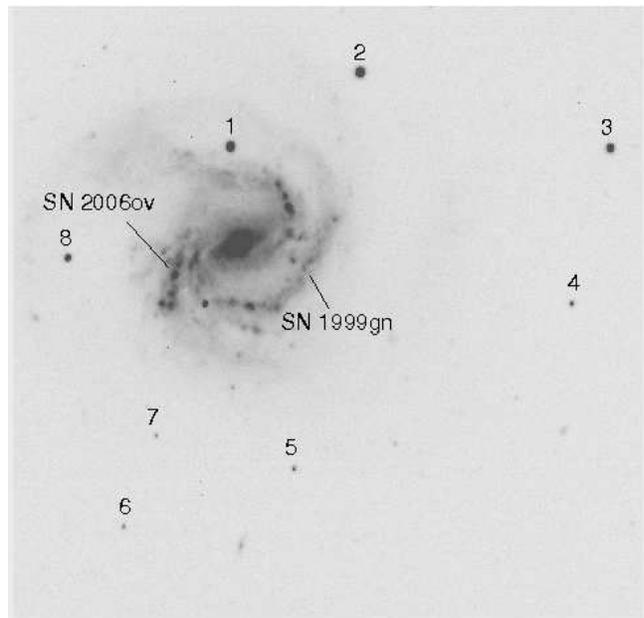}
\caption[SN 2006ov in NGC 4303]{SN 2006ov and position of SN 1999gn in NGC 4303 (V band image obtained on 2006 December 20 with the Copernico 1.82-m telescope of Mt. Ekar, Asiago (Italy)).The labels indicate the stars of the local sequence, as in Tab. \ref{ph_seq_06ov}. East is down, North is to the left.}
}
\end{center}
\end{figure}

SN 2006ov was discovered  by Nakano \& Itagaki (\citealp{Nakano06}) 
on 2006 November 24.86 UT with a 0.60-m reflector. The magnitude at discovery was 14.9.
This object was located at $\alpha
= 12^{h} 21^{m} 55^{s}.30$, $\delta$ = +04$\degr$ 29$\arcmin$ 16$\farcs7$ (equinox 2000.0), 
which is 5".5 East and 51" North of the center of the galaxy. 
Nothing was visible at this location on an 
exposure obtained on 2006 May 4 (to a limiting mag 19.5).
\cite{Blondin06} reported that a spectrogram (range 350-740 nm) of
2006ov  obtained by P. Berlind on Nov. 25.56 UT with the F.
L. Whipple Observatory 1.5-m telescope (+ FAST), showed that SN 2006ov was a
type-II supernova roughly one month past explosion.  Indeed the spectrum was similar
to that of the type IIP SN 2005cs at 36 days
past explosion.  Adopting a recessional velocity of 1570 km s$^{-1}$ for the host
galaxy (from ``The Updated Zwicky Catalog"),
the absorption minimum of the H${\alpha}$ line (rest 656.3 nm) indicates an expansion velocity of 4000 km s$^{-1}$ (\citealp{Blondin06}).

The adopted explosion epochs are JD = 2451520 $\pm$ 10 and JD = 2453974 $\pm$ 6 for SN 1999gn and SN 2006ov respectively, 
as inferred from spectroscopic and photometric informations (see section 4.2 and 5.1). Main data of SN 1999gn and SN 2006ov and their host galaxy are reported in Tab.\ref{datagal_gd}.

\begin{table*}
\begin{center}{
\caption{Main observational data of SN 1999gn, SN 2006ov, SN 2002gd, SN 2003Z, SN 2004eg  and their host galaxies. $m_{V}^{pl}$ is the mean plateau magnitude in the V band.} \label{datagal_gd}
\tiny
\begin{tabular}{ccccccccccc}\\ \hline \hline
SN Data   & 2002gd &&  2003Z & & 2004eg& &1999gn & & 2006ov &\\ \hline
$\alpha$  (J2000.0) &  23$^h$14$^m$36\fs98 & $\S$ &09h07m32\fs46 & $\spadesuit$ & 04h28m08\fs26 & $\diamond$ &12h21m57\fs02  & $\clubsuit$ & 12h21m55\fs30 & $\ominus$ \\
$\delta$ (J2000.0) &$+04\degr30\arcmin05\farcs7$ & $\S$ & $+60\degr29\arcmin17\farcs5$ & $\spadesuit$& $+21\degr39\arcmin18\farcs3$ & $\diamond$ &$+04\degr27\arcmin45\farcs6$ & $\clubsuit$&$+04\degr29\arcmin16\farcs7$ & $\ominus$\\
Offset SN-Gal.Nucleus &  36$^{\prime\prime}.8$E, 10$^{\prime\prime}$.8N & $\S$& 8$^{\prime\prime}$.4W, 31$^{\prime\prime}$.0N & $\spadesuit$& 20$^{\prime\prime}$W, 1$^{\prime\prime}$.4S & $\diamond$&31$^{\prime\prime}.7$E, 39$^{\prime\prime}.8$S & $\clubsuit$& 5$^{\prime\prime}.5$E, 51$^{\prime\prime}$N & $\ominus$  \\
Discovery Date (UT) &  2002 Oct 5.31 & $\S$&2003 Jan 29.7 & $\spadesuit$& 2004 Sept 1.488 & $\diamond$&1999 Dec 17.22 & $\clubsuit$&2006 Nov 24.86 & $\ominus$  \\
Discovery Julian Date &  2452552.81 & $\S$& 2452669.2 & $\spadesuit$& 2453249.5 & $\diamond$&2451529.7 & $\clubsuit$& 2454064 & $\ominus$\\
Adopted Explosion Epoch (JD) & 2452552 & $\times$&2452665 & $\times$& 2453170 & $\times$&2451520 & $\times$ & 2453974 & $\times$\\
Discovery Magnitude &  m=18.9 & $\S$& m=16.7 & $\spadesuit$ & m=19.5 & $\diamond$ &m=16.0 & $\clubsuit$& m=14.9 & $\ominus$\\
$m_{V}^{pl}$&  17.56 & $\times$& 17.53 & $\times$ & 19.16 & $\times$ &- & $\times$& 15.73 & $\times$\\
Total Extinction $A_{B,tot}$ &  0.243 &  $\times$& 0.141 &  $\times$& 1.635 &  $\times$ &0.081 &  $\times$& 0.081 &  $\times$  \\ \hline 
Host Galaxies Data &NGC 7537&&NGC 2742& &UGC 3053&& \multicolumn{3}{c}{NGC 4303}& \\ \hline
$\alpha$ (J2000.0) &23$^h$14$^m$34\fs50 & $\dag$& 09h07m33\fs53 & $\dag$& 04h28m09\fs6 & $\dag$&\multicolumn{3}{c}{12h21m54\fs9 \hspace{0.6cm} $\dag$}&\\
$\delta$ (J2000.0) &  $+04\degr29\arcmin54\farcs1$ & $\dag$&$\delta$  $+60\degr28\arcmin45\farcs6$ & $\dag$& $+21\degr39\arcmin19\farcs0$ & $\dag$&\multicolumn{3}{c}{$+04\degr28\arcmin25\farcs0$ \hspace{0.45cm} $\dag$}& \\
Morph. Type &SAbc: & $\dag$&SA(s)c: & $\dag$&Scd: & $\dag$&\multicolumn{3}{c}{ SAB(rs)bc \hspace{0.75cm}   $\dag$}&\\
B Magnitude & 13.86 & $\dag$& 12.03 & $\dag$ & 14.75 & $\dag$ &\multicolumn{3}{c}{10.18 \hspace{1.3cm}   $\dag$}&\\
Galactic Extinction $A_{B}$ &  0.243 & $\otimes$& 0.141 & $\otimes$& 1.635 & $\otimes$&\multicolumn{3}{c}{0.081\hspace{1.35cm}   $\otimes$}&\\
Diameters &   2'.2 x  0'.6 &$\dag$& 3'.0 x  1'.5 &$\dag$& 0'.9 x  0'.7 &$\dag$&\multicolumn{3}{c}{6'.5 x  5'.8\hspace{0.85cm} $\dag$}& \\
$v_{Vir}$ (km s$^{-1}$) & 2698 & $\star$& 1511 & $\star$& 2434 & $\star$& \multicolumn{3}{c}{1616\hspace{1.4cm}  $\star$}& \\
$\mu$ ($H_{0}$=72 km s$^{-1}$ Mpc$^{-1}$) &  32.87& $\times$&  31.7 & $\times$& 32.64& $\times$&\multicolumn{3}{c}{30.5 \hspace{1.4cm} $\times$}&\\
\hline\hline
\multicolumn{11}{c}{ $\clubsuit$ \cite{Dimai99}; $\times$ this paper; $\ominus$ \cite{Nakano06}; $\dag$ NED; $\otimes$ \cite{Schlafly11}; $\star$ HYPERLEDA;}\\
 \multicolumn{11}{c}{ $\S$ \cite{Klotz02}; $\spadesuit$ \cite{Boles03}; $\diamond$ \cite{Young04}.}\\

\end{tabular}
}
\end{center}
\end{table*}

\section{Summary of the observations}

In this paper we present new data for five type II-P SNe for the first time,
along with other data available in the literature.
Images and spectra of the 5 SNe were obtained using the following instruments:

- the 1.82-m Copernico Telescope with AFOSC (Asiago-Mt. Ekar, Italy); 

- the 0.8-m Teramo-Normale Telescope (TNT) with TK512CB1-1 CCD (Teramo, Italy);

- the ESO 3.6m with EFOSC2 (La Silla, Chile); 

- the ESO 3.58-m New Technology Telescope (NTT) with EMMI (La Silla, Chile);

- the 2.3-m telescope at the Siding Spring Observatory (SSO) with DBS (Australia); 

- the 74-inch  telescope of the Mount Stromlo Observatory (MSO) plus B\&C (Australia);
 
- the Calar Alto (CAHA) 2.2-m telescope  with CAFOS (Calar Alto Observatory, Spain);

- the 2.5-m Ir\'en\'ee du Pont telescope plus WFCCD (Las Campanas Observatory, Chile);

- the 1.0-m SWOPE telescope (Las Campanas Observatory,  Chile); 
 
- the 2.0-m Faulkes Telescope (FT) North with HawkCam (Haleakala, Hawaii, USA); 

- the 2.0-m MAGNUM with MIP (Haleakala, Hawaii, USA)\footnote{\it \cite{Yoshii02,Yoshii03}};

- the 2.0-m Liverpool Telescope (LT) plus RATCam (La Palma, Spain); 

- the 2.56-m Nordic Optical Telescope (NOT) plus ALFOSC (La Palma, Spain); 

- the 4.2-m William Herschel Telescope (WHT) equipped with ISIS (La Palma, Spain); 

- the 1.0m  Jacobus Kapteyn Telescope (JKT) with JAG (La Palma, Spain);

- the 3.58-m Telescopio Nazionale Galileo (TNG) with Dolores (La Palma, Spain); 

- the 0.41-m Newton Telescope (Nw 0.4m) of the Gruppo Astrofili di Padova (GAP) 
equipped with an Apogee AP47p CCD Camera (Padova, Italy).

After the usual initial (bias, overscan, normalized flat-fields) corrections,
photometric data were reduced following standard prescriptions (see e.g. \citealp{Pastorello07}) 
using tasks developed by the Asiago-Padova SN Group 
in the IRAF environment, while the spectroscopic data were reduced using traditional IRAF tasks.
Two different methods were used to obtain the SN magnitudes: template subtraction and
 point spread function (PSF) - fitting technique, depending on the characteristics of the background where the supernova 
exploded and/or the availability of pre-explosion images of the SN site in the given bands. 
The explosion epochs of all SNe of our sample were estimated through spectroscopic and 
photometric comparisons with the well studied SN 2005cs for which the explosion 
date was well constrained (about 0.5 days, \citealp{Pastorello09}). The underlying assumption is that low luminosity type IIP SNe have intrinsically similar 
properties (see section \ref{sec:colour}).

\section{Light curves}

Photometric data for SN 2002gd, SN 2003Z, SN 2004eg and SN 2006ov
are presented here but 
we regret that no photometric follow up was performed for SN 1999gn.

\subsection{Individual objects}

\subsubsection{SN 2002gd}
The contribution of a large number of facilities allowed an extensive observational 
campaign for this SN during 
the photospheric phase in the optical bands. These are complemented by few 
near-infrared (NIR) data in the late plateau phase calibrated using the 2MASS catalogue (\citealp{Skrutskie06}).
Optical and NIR magnitudes of SN~2002gd are reported in
Tab. \ref{sn_mag2}  
while the magnitudes of the local sequence stars (with associated errors) 
are in Tab. \ref{ph_seq2}.

After the steep rise observed during the first few days after discovery,
 the SN reached the maximum in the optical bands around October 13, 2002. 
 After the maximum the luminosity declined slightly in all bands (especially in B) for about 20 days. Then the SN entered the plateau phase with a constant luminosity.
The plateau of SN~2002gd lasted about three months, although a further minor rebrightening
was observed at $\sim$ 70 days. The 
U band light curve had a different behaviour, showing a monotonic
decline.

At about 110 days after the explosion, the SN luminosity dropped steeply,
 that usually takes the light curve to the  radioactive tail. 
Unfortunately this transition was observed only in the initial phase
because SN~2002gd was lost behind the Sun. 
We tried to recover the SN during the nebular phase
but the SN was not visible  and only upper limits in the SN luminosity
were obtained.

These limits allow us to estimate an upper limit for the  $^{56}$Ni mass ejected in the explosion of SN 2002gd, 
that turned out to be very small, $\leq$ 0.003M$_{\odot}$. 
Missing a real detection and, therefore, an estimate of the late--time 
magnitude decline rate, we cannot exclude
a dimming of the luminosity due to dust formation into the ejecta that could lead us 
to underestimate the upper limit of the mass of $^{56}$Ni.

\begin{figure}
\begin{center}
\includegraphics[width=6.5cm, angle=270]{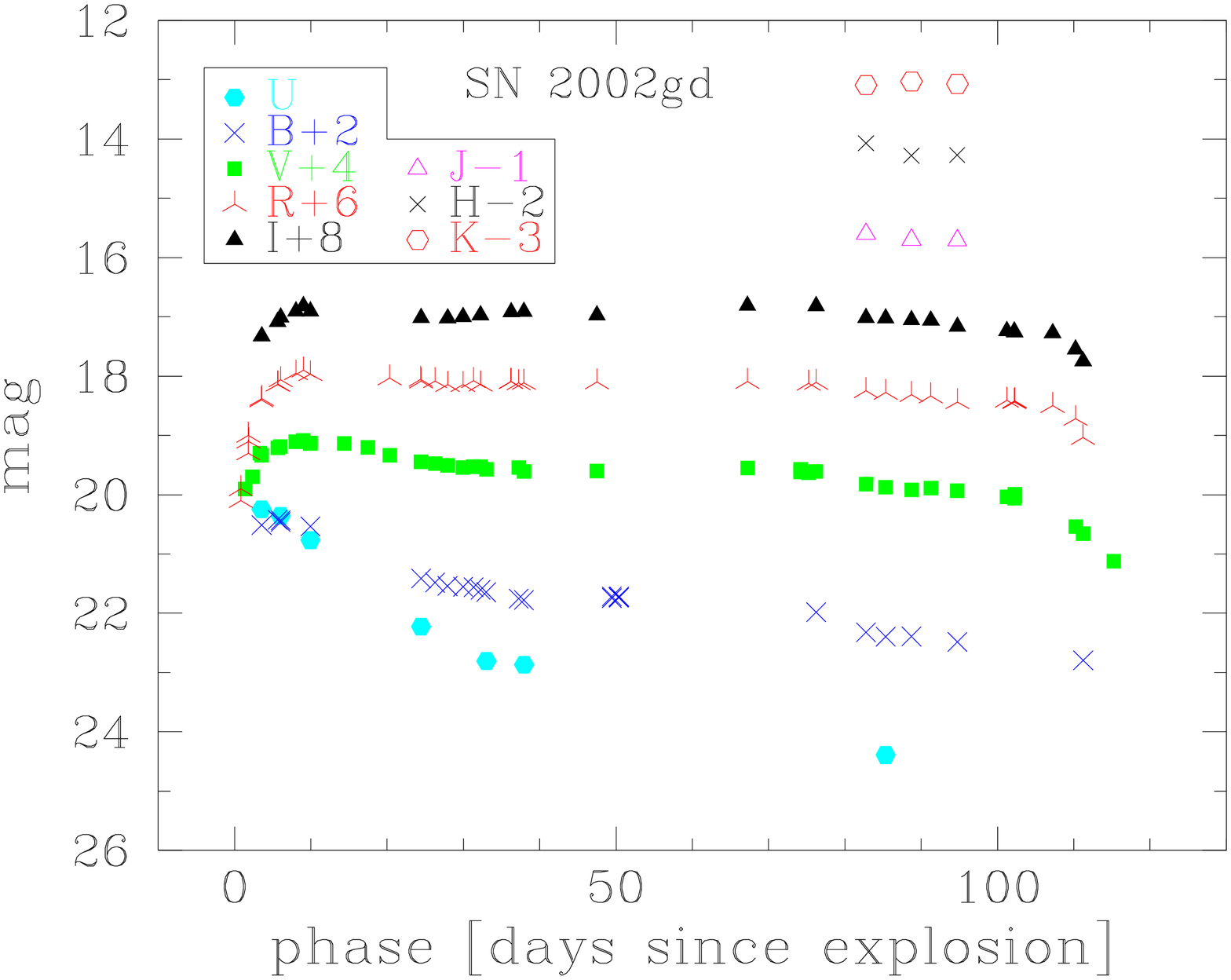}
\caption[Early-time UBVRI light curves of SN~2002gd]{Early-time UBVRIJHK light curves of SN~2002gd. 

} \label{sn02gd_lc_fin}
\end{center}
\end{figure}

\begin{table*}
\begin{center}{

\caption[New photometric data of moderately faint SNe II--P]{Photometry of SN 2002gd.} \label{sn_mag2}
\tiny
\begin{tabular}{p{0.95cm}ccccccccccc} \\ \hline
Date & JD (2400000+) & U & B & V & R & I & J & H & K & Instr. \\ \hline \hline  
\multicolumn{11}{c}{SN 2002gd}\\  \hline
08/10/02 & 52555.52 & -- & -- & -- & 17.40 (0.01) & -- & --&--&--&1 \\
08/10/02 & 52555.55 & 16.25 (0.01) & 17.51 (0.01) & 17.33 (0.01) & 17.38 (0.01) & 17.33 (0.02) &--&--&--& 1 \\
10/10/02 & 52557.66 & -- & 17.43 (0.01) & 17.21 (0.01) & 17.14 (0.01) & 17.08 (0.01) & --&--&--&2 \\
10/10/02 & 52557.67 & -- & -- & -- & 17.14 (0.01) & -- & --&--&--&2 \\
10/10/02 & 52558.04 & 16.36 (0.03) & 17.43 (0.01) & 17.19 (0.02) & 17.07 (0.02) & 17.01 (0.03) & --&--&--&3 \\
10/10/02 & 52558.04 & -- & 17.46 (0.01) & -- & -- & -- & --&--&--&3 \\
12/10/02 & 52560.04 & -- & -- & 17.10 (0.02) & 16.95 (0.01) & 16.90 (0.01) &--&--&--& 3 \\
13/10/02 & 52561.02 & -- & -- & 17.08 (0.07) & 16.90 (0.13) & 16.82 (0.09) & --&--&--&3 \\
14/10/02 & 52561.93 & -- & -- & 17.14 (0.01) & -- & -- & --&--&--&3 \\
14/10/02 & 52561.95 & 16.77 (0.02) & 17.54 (0.02) & 17.13 (0.01) & 16.96 (0.01) & 16.91 (0.01) & --&--&--&3 \\
18/10/02 & 52566.41 & -- & -- & 17.13 (0.11) & -- & -- &--&--&--& 4 \\
24/10/02 & 52572.34 & -- & -- & 17.34 (0.06) & 17.03 (.06) & -- &--&--&--& 5 \\
28/10/02 & 52576.38 & -- & -- & -- & 17.05 (0.12) & -- & --&--&--&4 \\
28/10/02 & 52576.44 & 18.23 (0.05) & 18.42 (0.02) & 17.45 (0.02) & 17.08 (0.02) & 17.02 (0.02) & --&--&--&6 \\ 
30/10/02 & 52578.30 & -- & 18.48 (0.12) & 17.48 (0.08) & 17.08 (0.09) & --  &--&--&--& 5 \\
01/11/02 & 52579.92 & -- & 18.54 (0.03) & 17.51 (0.02) & 17.14 (0.02) & 17.02 (0.02) & --&--&--&7 \\
03/11/02 & 52581.93 & -- & 18.55 (0.11) & 17.54 (0.03) & 17.14 (0.02) & 17.00 (0.02) & --&--&--&7 \\
04/11/02 & 52583.31 & -- & 18.56 (0.15) & 17.53 (0.03) & 17.08 (0.05) & -- & --&--&--&5 \\
05/11/02 & 52584.23 & -- & 18.60 (0.04) & 17.53 (0.02) & 17.14 (0.02) & 16.97 (0.04) & --&--&--&6 \\
06/11/02 & 52585.01 & 18.81 (0.17) & 18.65 (0.04) & 17.57 (0.04) & -- & -- & --&--&--&3 \\ 
09/11/02 & 52588.23 & -- & -- & -- & 17.09 (0.20) & -- &--&--&--& 5 \\
09/11/02 & 52588.26 & -- & -- & -- & 17.09 (0.10) & 16.92 (0.08) & --&--&--&5 \\       
10/11/02 & 52589.26 & -- & 18.75 (0.20) & 17.54 (0.11) & 17.11 (0.04) & -- &--&--&--& 4 \\
11/11/02 & 52589.94 & 18.87 (0.33) & 18.78 (0.04) & 17.61 (0.03) & 17.11 (0.02) & 16.91 (0.02) & --&--&--&3 \\
20/11/02 & 52599.45 & -- & -- & 17.60 (0.04) & 17.10 (0.04) & 16.93 (0.10) & --&--&--&5 \\
22/11/02 & 52601.44 & -- & 18.71 (0.08) & -- &  -- &  -- & --&--&--&8 \\
22/11/02 & 52601.45 & -- & 18.74 (0.08) & -- &  -- &  -- & --&--&--&8 \\
23/11/02 & 52602.34 & -- & 18.72 (0.10) & -- &  -- &  -- & --&--&--&8 \\
23/11/02 & 52602.35 & -- & 18.74 (0.13) & -- &  -- &  -- & --&--&--&8 \\
23/11/02 & 52602.44 & -- & 18.72 (0.28) & -- &  -- &  -- & --&--&--&8 \\  
10/12/02 & 52619.25 & -- & -- & 17.55 (0.12) & 17.09 (0.04) & 16.81 (0.02) & --&--&--&4 \\
17/12/02 & 52626.20 & -- & -- & 17.60 (0.05) & -- & -- & --&--&--&5 \\
18/12/02 & 52627.25 & -- & -- & 17.63 (0.09) & 17.12 (0.03) & -- & --&--&--&4 \\ 
19/12/02 & 52628.23 & -- & 19.06 (0.10) & 17.65 (0.05) & 17.12 (0.05) & 16.83 (0.17) & --&--&--&5 \\
26/12/02 & 52634.77 & -- & 19.55 (0.03) & 17.84 (0.02) & 17.28 (0.02) & 17.01 (0.02) & 16.60 (0.04) & 16.07 (0.05) & 16.09 (0.04) &9 \\ 
28/12/02 & 52637.35 & 20.85 (0.30) & 19.57 (0.04) & 17.86 (0.02) & 17.29 (0.02) & 17.04 (0.04) & --&--&--&10 \\ 
01/01/03 & 52640.74 & -- & 19.64 (0.03) & 17.92 (0.02) & 17.35 (0.02) & 17.11 (0.02) & 16.70 (0.06) & 16.28 (0.04) & 16.03 (0.08)&9 \\ 
03/01/03 & 52643.28 & -- & -- & 17.95 (0.05) & 17.39 (0.05) & 17.13 (0.05) & --&--&--&6 \\
07/01/03 & 52646.74 & -- & 19.70 (0.03) & 18.03 (0.03) & 17.48 (0.02) & 17.18 (0.02) & 16.71 (0.04) & 16.27 (0.06) & 16.07 (0.07) &9 \\
13/01/03 & 52653.24 & -- & -- & 18.04 (0.14) & 17.41 (0.12) & 17.24 (0.30) & --&--&--&5 \\
14/01/03 & 52654.21 & -- & -- & 18.06 (0.03) & 17.44 (0.03) & 17.24 (0.03) & --&--&--&6 \\
14/01/03 & 52654.27 & -- & -- & 17.99 (0.08) & 17.42 (0.08) & 17.26 (0.13) & --&--&--&5 \\
14/01/03 & 52654.28 & -- & -- & 18.03 (0.11) & 17.42 (0.15) & -- & --&--&--&6 \\
19/01/03 & 52659.23 & -- & -- & -- & 17.50 (0.05) & 17.28 (0.03) & --&--&--&4 \\
22/01/03 & 52662.25 & -- & -- & 18.54 (0.11) & 17.72 (0.20) & 17.55 (0.07) & --&--&--&5 \\
23/01/03 & 52663.24 & -- & 19.79 (0.09) & 18.66 (0.06) & 18.04 (0.03) & 17.75 (0.02) & --&--&--&6 \\
27/01/03 & 52667.25 & -- & -- & 19.12 (0.30) & -- & -- & --&--&--&5 \\ 
03/06/03 & 52794.09 &  --&--&--&--&--&$\geq$ 19.8 & $\geq$ 19.2 & $\geq$ 18.1 & 9 \\
04/06/03 & 52795.07 & -- & $\geq$ 22.7 & $\geq$ 22.6 & $\geq$ 22.8 & $\geq$ 22.0 &--&--&--& 9 \\
08/06/03 & 52799.09 & --&--&--&--&--& $\geq$ 19.8 & -- & -- & 9 \\
23/08/03 & 52874.54 & -- & -- & -- & $\geq$ 22.90 & $\geq$ 22.67 &--&--&--& 6 \\ \hline \hline

\multicolumn{11}{c}{1 = ESO3.6m + EFOSC2; 2 = NTT + EMMI; 3 = SSO2.3m; 4 = Padova-Nw0.4m + CCD; 5 = TNT + CCD; 6 = Asiago1.82m + AFOSC;}\\
\multicolumn{11}{c}{7 = SSO-40inch + WFI;  8 = JKT + JAG; 9 = MAGNUM 2m + MIP; 10 = CAHA2.2m + CAFOS.}\\

\end{tabular}

}
\end{center}
\end{table*}

 \begin{table}
\begin{center}{
\caption[Optical magnitudes of the sequence stars in the fields of the
SNe]{Magnitudes of the local sequence stars in the fields of 
SN 2002gd.  } \label{ph_seq2}
\tiny
\begin{tabular}{p{0.05cm}ccccc}\\ \hline 
Star & U & B & V & R & I  \\ \hline \hline  
\multicolumn{6}{c}{SN 2002gd} \\ \hline
1 & 17.39 (0.05) & 17.07 (0.02) & 16.00 (0.02) & 15.54 (0.02) & 15.09 (0.01)  \\
2 & -- & 20.67 (0.05) & 19.15 (0.03) & 18.24 (0.02) & 17.29 (0.02)  \\
3 & -- & 21.40 (0.09) & 19.80 (0.05) & 18.65 (0.03) & 17.55 (0.03)  \\
4 & -- & 19.26 (0.04) & 17.67 (0.02) & 16.80 (0.02) & 15.98 (0.01)  \\
5 & 18.35 (0.10) & 17.43 (0.02) & 16.05 (0.01) & 15.35 (0.01) & 14.71 (0.02)  \\
6 & -- & 21.63 (0.07) & 19.83 (0.04) & 18.72 (0.03) & 17.48 (0.02)  \\
7 & -- & 20.39 (0.05) & 19.66 (0.05) & 19.36 (0.05) & 18.86 (0.04) \\
8 & 17.54 (0.05) & 17.67 (0.02) & 16.85 (0.02) & 16.52 (0.01) & 16.18 (0.01)  \\
9 & 19.03 (0.20) & 19.73 (0.04) & 19.26 (0.03) & 19.18 (0.04) & 18.94 (0.03) \\ \hline \hline 
\end{tabular}
}
\end{center}
\end{table}

\subsubsection{SN 2003Z}

\begin{figure}
\begin{center}
\includegraphics[width=6.5cm, angle=270]{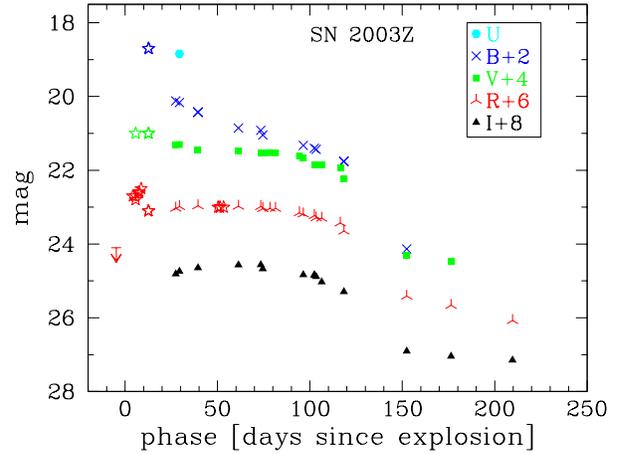}
\caption[BVRI light curves of SN 2003Z]{UBVRI light curves of SN 2003Z.  The unfiltered prediscovery
limit is
shown with respect to the R magnitude scale. Open star symbols are filtered magnitudes from http://www.rochesterastronomy.org
(measured by Bruce L. Gary), all other symbols represent magnitudes published in this paper.} \label{fig:sn03Z_lc}
\end{center}
\end{figure}

Our photometric follow-up observations started about 1 month after the adopted 
explosion epoch and lasted until day 210, in the nebular phase.
PSF-fitting photometric measurements of SN~2003Z are reported in Tab. \ref{photometry_03Z}, 
while the magnitudes of the sequence stars (Fig. \ref{sn03Z_field}) are listed in Tab. \ref{ph_seq_03Z}.
Early-time photometry (both filtered and unfiltered) is adopted from the 
Bright Supernova web 
site\footnote{\it http://www.rochesterastronomy.org}. 
These data give useful constraints on the
photometric evolution of the SN during the first month after the explosion and
show that SN~2003Z was discovered when it was very young, likely a few days after the shock breakout.
Merging these unfiltered measurements with those presented in this paper,
allows us to infer
a plateau duration of about 100 days (see Fig. \ref{fig:sn03Z_lc}). 
The end of the plateau is well determined, after which we observe a 
large luminosity drop lasting a few weeks, until the SN light curve 
reaches the exponential tail. Despite the relatively
large uncertainty in the late--time photometry, the average luminosity decline 
is rather close to that expected for the luminosity decline of $^{56}$Co 
into $^{56}$Fe (0.98 mag/100days, see Tab. \ref{slope_sn03Z}).

\begin{table*}
\begin{center}{
\caption[New photometry of extremely faint SNe II--P]{Photometry in the optical bands of SN 2003Z.
The data have been obtained using several telescopes in Italy and Canary Islands
(Spain).} \label{photometry_03Z}
\tiny
\begin{tabular}{p{0.95cm}ccccccc}\\ \hline
Date & JD (2400000+) & U & B & V & R & I & Ins. \\ \hline \hline  
\multicolumn{8}{c}{SN 2003Z} \\ \hline
21/02/03 & 52692.40 & -- & -- & -- & 17.03 (0.05) & -- & 1  \\
21/02/03 & 52692.44 & -- & 18.12 (0.02) & 17.33 (0.03) & -- &  16.81 (0.02) & 1 \\  
23/02/03 & 52694.41 & 18.84 (0.06) & 18.16 (0.05) & 17.31 (0.04) & 16.98 (0.02) & 16.74 (0.02) & 2 \\
05/03/03 & 52704.40 & -- & 18.43 (0.06) & -- & -- & -- & 2 \\
05/03/03 & 52704.41 & -- & 18.43 (0.12) & -- & -- & -- & 2 \\
05/03/03 & 52704.42 & -- & 18.42 (0.10) & -- & -- & -- & 2 \\
05/03/03 & 52704.43 & -- & 18.43 (0.07) & 17.44 (0.04) & 16.96 (0.04) & 16.65 (0.02) & 2 \\
27/03/03 & 52726.32 & -- & 18.86 (0.07) & 17.48 (0.02) & 16.97 (0.03) & 16.57 (0.02) & 2 \\
08/04/03 & 52738.41 & -- & 18.92 (0.15) & 17.52 (0.05) & 16.98 (0.03) & 16.56 (0.03) & 2 \\ 
10/04/03 & 52739.50 & -- & 19.04 (0.05) & 17.52 (0.01) & 17.04 (0.01) & 16.67 (0.01) & 1 \\
13/04/03 & 52743.40 & -- & -- & 17.52 (0.11) & 17.03 (0.07) & -- & 3 \\  
16/04/03 & 52746.36 & -- & -- & 17.53 (0.16) & 17.03 (0.07) & -- & 3 \\   
29/04/03 & 52759.36 & -- & -- & 17.61 (0.07) & 17.15 (0.08) & -- & 3 \\ 
01/05/03 & 52761.46 & -- & 19.33 (0.08) & 17.67 (0.03) & 17.19 (0.02) & 16.83 (0.02) & 1 \\
07/05/03 & 52767.43 & -- & 19.40 (0.26) & 17.86 (0.10) & 17.23 (0.09) & 16.83 (0.14) & 2 \\
07/05/03 & 52767.45 & -- & -- & -- & -- & 16.85 (0.09) & 2 \\
08/05/03 & 52768.42 & -- & 19.43 (0.06) & 17.87 (0.04) & 17.29 (0.03) & 16.89 (0.02) & 2 \\
11/05/03 & 52771.38 & -- & -- & 17.86 (0.08) & 17.30 (0.14) & 17.03 (0.06) & 3 \\ 
21/05/03 & 52781.46 & -- & -- & 17.94 (0.07) & 17.43 (0.07) & -- & 3 \\
23/05/03 & 52783.38 & -- & 19.75 (0.23) & 18.24 (0.07) & -- & -- & 2 \\
23/05/03 & 52783.40 & -- & 19.76 (0.25) & -- & 17.65 (0.13) & 17.29 (0.09) & 2 \\ 
26/06/03 & 52817.41 & -- & 22.13 (0.28) & 20.31 (0.11) & 19.41 (0.05) & 18.90 (0.04) & 1 \\
21/07/03 & 52841.36 & -- & -- & 20.47 (0.32) & 19.66 (0.23) & 19.04 (0.15) & 2 \\
23/08/03 & 52874.63 & -- & -- & -- & 20.07 (0.15) & 19.15 (0.21) & 2 \\ \hline \hline 
\end{tabular}
\\
1 = TNG + Dolores; 2 = Asiago 1.82m + AFOSC; 3 = Nw 0.4m + CCD\\
}
\end{center}
\end{table*}

\begin{table}
\begin{center}{
\caption[Magnitudes of the sequence stars in the fields of SN 2003Z]{Magnitudes of the stars of the local sequences of SN 2003Z.  
The numbers in brackets are the r.m.s. of the
available measurements. If no error is reported, only 
a single estimate is available.} \label{ph_seq_03Z}
\tiny
\begin{tabular}{cccccc}\\ \hline 
Star & U & B & V & R & I  \\ \hline \hline  
\multicolumn{6}{c}{SN 2003Z} \\ \hline
1 & -- & 18.87 (0.03) & 17.65 (0.01) & 16.89 (0.02) & 16.26 (0.01)  \\ 
2 & 18.08 (--) & 17.35 (0.01) & 16.43 (0.01) & 16.00 (0.01) & 15.54 (0.01)  \\
3 & -- & 19.03 (0.01) & 17.53 (0.01) & 16.59 (0.01) & 15.62 (0.01)  \\
4 & 14.13 (--) & 13.88 (--) & 13.29 (0.03) & 12.93 (0.02) & --   \\
5 & 14.32 (--) & 14.04 (--) & 13.41 (0.03) & 13.05 (0.02) & --   \\
6 & -- & 17.44 (0.02) & 16.79 (0.01) & 16.41 (0.01) & 16.06 (0.01) \\
7 & -- & 18.19 (0.02) & 16.98 (0.01) & 16.13 (0.01) & 15.37 (0.01) \\
8 & -- & -- & 16.84 (0.12) & 16.14 (0.01) & --  \\
9 & -- & 17.76 (--) & 16.63 (0.05) & 15.89 (0.02) & 15.06 (--)  \\
10 & 19.75 (--) & 18.40 (0.02) & 17.28 (0.02) & 16.52 (0.01) & 15.73 (0.02)   \\
11 & -- &  -- & 13.41 (--) & 13.07 (0.02) & --  \\
12 & -- &  15.71 (--) & 14.99 (--) & 14.69 (0.03) & 14.20 (--)  \\
13 & -- &  -- & 16.85 (--) & 16.14 (0.05) & --  \\
14 & 15.95 (--) &  -- & 14.68 (--) & 14.22 (0.02) & 13.69 (0.01)  \\
15 & -- &  -- & 15.62 (--) & 15.25 (0.04) & --  \\
16 & -- & 18.72 (0.09) & 18.14 (0.01) & 17.82 (0.01) & 17.39 (0.01)  \\
17 & -- & 18.82 (0.02) & 18.18 (0.01) & 17.80 (0.01) & 17.41 (0.01)  \\
18 & -- & 19.67 (0.02) & 18.29 (0.01) & 17.37 (0.01) & 16.60 (0.01)  \\ \hline\hline
\end{tabular}
}
\end{center}
\end{table}

\begin{table}
\begin{center}{
\caption[Slopes of the light curves of SN 2003Z in the B, V, R, I bands]
{Slopes of the light curves of SN 2003Z in the B, V, R, I bands
(in mag/100$^{d}$). The phase is relative to the explosion date (JD = 2452665).} \label{slope_sn03Z}
\begin{tabular}{cccccc} \\ \hline
band & 10--40$^d$ & 35--100$^d$ & 90--110$^d$ & 115--155$^d$ & $\ge$150$^d$\\ \hline \hline  
$\gamma_B$ & 6.32$^{\ast}$ & 1.60 & 1.44 & 6.99 & -- \\
$\gamma_V$ & 1.73 & 0.36 & 2.32 & 6.37 & 0.67 \\
$\gamma_R$ & -0.54 & 0.32 & 1.21 & 5.37 & 1.05 \\
$\gamma_I$ & -1.20 & 0.31 & 1.68 & 4.73 & 0.58 \\ \hline\hline 
\end{tabular}}
\end{center}
\begin{center}
$^{\ast}$ the measured slope of the B band light curve changes \\
from $\gamma_B$ = 9.65 (10--30 days) to $\gamma_B$ = 2.58 (25--40 days).
\end{center}
\end{table}

\subsubsection{SN 2004eg}

Our optical photometry in the g'BVRI bands, collected
 in Tab.  \ref{photometry_04eg}, span about 4 months.  
The sloan r' and i' magnitudes collected with the SWOPE telescope were converted 
into Johnson-Bessel R and I magnitudes making use of the relation presented in 
\cite{Smith02}:

\begin{equation}
V-R = 0.59 (g'-r') + 0.11
\end{equation}
and 

\begin{equation}
R-I  = 1.00 (r'-i') + 0.21 (for (r'-i') < 0.95)
\end{equation}

\begin{figure}
\begin{center}
\includegraphics[width=6.5cm, angle=270]{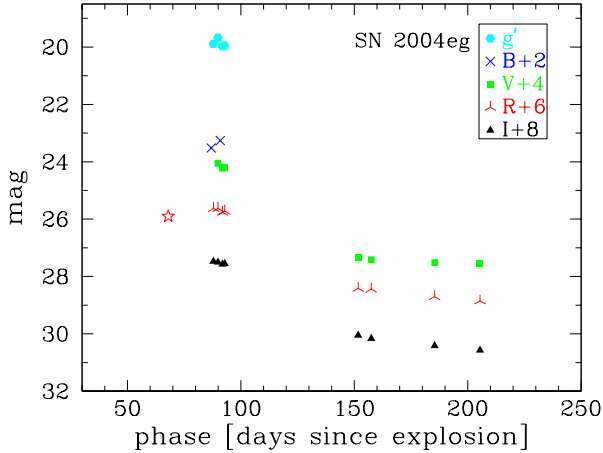}
\caption[BVRI light curves of SN 2004eg]{g'BVRI light curves of SN 2004eg. The open symbol is from \cite{Young04}, filled ones are magnitudes published in this paper.} \label{sn04eg_lc}
\end{center}
\end{figure}

 Photometric measurements of SN 2004eg were performed using a PSF-fitting method.
Photometric errors were estimated with artificial star experiments: the large errors are 
consistent with the faint SN magnitudes. For calibration we used the local stellar sequence
shown in Fig. \ref{sn04eg_field}. Their magnitudes are reported in Tab \ref{ph_seq_04eg}.
The poorly sampled light curve of SN 2004eg is shown in Fig. \ref{sn04eg_lc}. A few photometric points
were collected in the plateau phase, and these allow to estimate the
SN luminosity in the recombination phase. Additionally, a few photometric
observations of the radioactive tail allow us to estimate the $^{56}$Ni mass (see Sect. \ref{sec:bolom}).

\begin{table*}
\begin{center}{
\caption[New photometry of extremely faint SNe II--P]{Photometry in the optical bands of SN 2004eg.
The data of this SN  have been obtained using TNG,SWOPE and  Asiago 1.82m telescopes.} \label{photometry_04eg}
\tiny
\begin{tabular}{p{0.95cm}cccccccc}\\ \hline
Date & JD (2400000+)& g' & B & V & R & I & Ins. \\ \hline \hline  
\multicolumn{6}{c}{SN 2004eg} \\ \hline
08/09/04  & 53256.91 & - & 21.02  (0.09) & - & - & - & 1 \\ 
09/09/04  & 53257.89 & 19.89  (0.03) & - & - &    18.13  (0.03) &   17.47   (0.03) & 1 \\
11/09/04  & 53259.89 & 19.68  (0.04) & - & 19.06  (0.04) & 18.13  (0.03) &   17.50   (0.03) & 1\\
12/09/04  & 53260.89 & - &  20.76  (0.07) & - & - & - & 1\\
13/09/04  & 53261.89 & 19.97  (0.04) & - & 19.22  (0.04) & 18.26  (0.04) &   17.57   (0.04) & 1\\
14/09/04  & 53262.85 & 19.95  (0.04) & - & 19.19  (0.04) & 18.22  (0.04) &   17.55   (0.04) & 1\\
11/11/04  & 53321.69 & - & - &  22.35  (0.33) &  20.91  (0.09) &  20.05   (0.10) &    2\\
16/11/04  & 53327.48 & - & - &  22.42  (0.25) &  20.94  (0.20) &  20.16   (0.32) &    3\\
15/12/04  & 53355.43 & - & - &  22.52  (0.36) &  21.20  (0.13) &  20.41   (0.11) &    3\\
03/01/05  & 53375.42 & - & - &  22.53  (0.36) &  21.35  (0.22) &  20.57   (0.19) &    3\\ \hline \hline

\end{tabular}
\\
1 = SWOPE 2 = TNG + Dolores; 3 = Asiago 1.82m + AFOSC\\
}
\end{center}
\end{table*}

\begin{table}
\begin{center}{
\caption[Magnitudes of the sequence stars in the fields of SN 2004eg]{Magnitudes of the stars of the local sequence.  
The numbers in brackets are the r.m.s. of the
available measurements. If no error is reported, only 
a single estimate is available.} \label{ph_seq_04eg}
\tiny
\begin{tabular}{cccccc}\\ \hline 
Star & g' & B & V & R & I  \\ \hline \hline  
\multicolumn{6}{c}{SN 2004eg} \\ \hline
1 &  18.00 (0.01) & 18.40 (0.03) & 17.69 (0.02) & 17.14 (0.02) & 16.62 (0.02)  \\
2 &  18.35 (0.01) & 18.92 (0.01) & 17.88 (0.02) & 17.23 (0.03) & 16.58 (0.01)  \\
3 &  - & - & 18.42 (0.02) & 17.79 (0.03) & 17.27 (0.01)\\
4 &  18.59 (0.01) & 19.02 (0.05) & 18.27 (0.01) & 17.72 (0.02) & 17.20 (0.01)  \\
5 &  18.13 (0.01) & 18.50 (0.04) & 17.85 (0.02) & 17.34 (0.02) & 16.82 (0.01)  \\
6 &  - & - & 16.49 (0.02) & 15.90 (0.01) & 15.31 (0.06) \\
7 & - & - & 15.72 (0.02) & 15.21 (0.03) & 14.70 (-) \\
8 & - & - & 18.22 (0.01) & 17.62 (0.03) & 17.04 (0.01) \\
9 & - & - & 17.48 (0.01) & 16.98 (0.04) & 16.45 (0.01) \\
10 & - & - & 15.82 (0.02) & 15.05 (0.03) & -\\
11 & - & - & 18.15 (0.01) & 17.29 (0.01) & 16.66 (-) \\
12 & - & - & 18.34 (0.05) & 17.44 (0.06) & 16.65 (-) \\
13 & - & - & 17.29 (0.05) & 16.51 (0.06) & 15.88 (-) \\
14 & - & - & 16.75 (0.04) & 16.21 (0.07) & 15.73 (-) \\\hline \hline

\end{tabular}
}
\end{center}
\end{table}

\subsubsection{SN 2006ov}

\begin{figure}
\begin{center}
\includegraphics[width=6.5cm, angle=270]{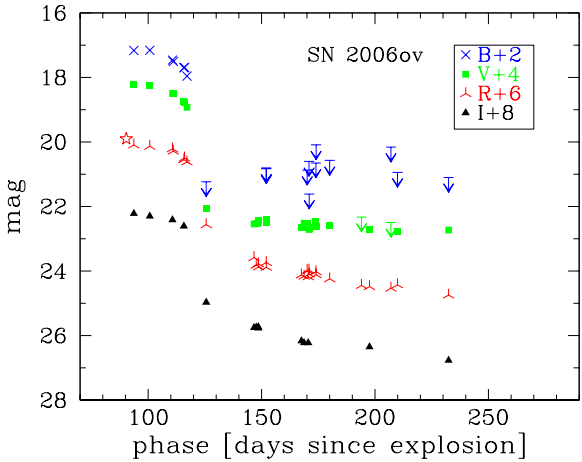}
\caption[BVRI light curves of SN 2006ov]{BVRI light curves of SN 2006ov. The open symbol is the unfiltered discovery magnitude (\citealp{Nakano06}) showed with respect to the R band. Filled ones are 
magnitudes published in this paper.}
 \label{sn06ov_lc}
\end{center}
\end{figure}

Photometric data were collected at 24 different epochs over a period of 150 days, starting from the end of the photospheric phase.
SN magnitudes are reported in Tab. \ref{photometry_06ov}, while those of the local 
sequence stars used for the calibration are collected 
in Tab. \ref{ph_seq_06ov}. The light curve of SN 2006ov is shown in Fig. \ref{sn06ov_lc}.

SN 2006ov exploded in a crowded  background region of the host galaxy. 
As long as the source was bright, the SN data were obtained with the PSF-fitting technique.
When the SN luminosity faded,  the photometric measurements were performed using the template 
subtraction technique. The errors were estimated by putting artificial stars with the same magnitude as the SN, in a number of locations close to the SN position,
and were estimated as the r.m.s. of the recovered magnitudes. 

Unfortunately, this SN was discovered near the end of the plateau phase.
A few points showed that in this phase the SN had a roughly constant magnitudes, 
about $m_{V} \simeq$ 15.7 mag. Between day $\sim$ 110 and day $\sim$ 140, after the end of the recombination phase, 
a steep decline of more than four magnitudes was observed in all bands. 
A slower decline is then observable beyond 150 days. 
In this phase, the SN becomes too faint to be observed in the B band and  only upper limits
were measured. Nevertheless, the luminosity decline at late epochs is still consistent
with the radioactive decay of $^{56}$Co into $^{56}$Fe (Sect. \ref{sec:bolom}).

\begin{table*}
\begin{center}{
\caption[New photometry of extremely faint SNe II--P]{New photometry in the optical bands for SN 2006ov.}
 \label{photometry_06ov}
\tiny
\begin{tabular}{p{0.95cm}ccccccc}\\ \hline
Date & JD (2400000+)  & B & V & R & I & Ins. \\ \hline \hline  
\multicolumn{7}{c}{SN 2006ov} \\ \hline
24/11/06 & 54064.34  & --      & --     &  14.90 (-) &     --   &   6  \\
28/11/06 & 54067.70  & 17.16 (0.08) & 15.73 (0.03) &  15.09 (0.02) &   14.72 (0.02) & 1\\
05/12/06 & 54074.73  & 17.16 (0.03) & 15.74 (0.02) &  15.13 (0.01) &   14.80 (0.01) & 2\\
15/12/06 & 54084.71  & 17.46 (0.05) & 16.00 (0.02) &  15.21 (0.01) &   14.91 (0.01) & 2\\
15/12/06 & 54085.11  & 17.50 (0.04) & 15.99 (0.02) &  15.28 (0.02) &     --   &    4\\
20/12/06 & 54089.72  & 17.70 (0.14) & 16.24 (0.04) &  15.55 (0.03) &   15.11 (0.03) & 1\\
20/12/06 & 54090.04  & 17.69 (0.05) & 16.29 (0.01) &  15.49 (0.01) &     --   &    4\\
21/12/06 & 54091.11  & 17.96 (0.02) & 16.43 (0.01) &  15.62 (0.01) &     --   &    4\\
30/12/06 & 54099.62  & $\geq$ 21.23  & 19.55 (0.15) &  17.56 (0.05) &   17.47 (0.11) & 2\\
20/01/07 & 54120.67  &  --     & 20.04 (0.06) &  18.58 (0.02) &   18.25 (0.07) &   2\\
21/01/07 & 54121.67  &  --     & 20.04 (0.18) &  18.82 (0.06) &   18.24 (0.07) &   2\\
22/01/07 & 54122.56  &  --     & 20.01 (0.10) &  18.77 (0.07) &   18.23 (0.07) &   2\\ 
22/01/07 & 54122.67  &  --     & 19.93 (0.23) &  18.87 (0.02) &   18.27 (0.10) &   2\\
25/01/07 & 54126.05  & $\geq$ 20.81  & 19.99 (0.21) &  18.73 (0.04) &    --   &    4\\  
25/01/07 & 54126.11  & $\geq$ 20.89  & 19.89 (0.22) &  18.88 (0.07) &    --   &    4\\
10/02/07 & 54141.56  &   --    & 20.15 (0.26) &  19.12 (0.02) &   18.66 (0.07) &   2\\
11/02/07 & 54142.76  &   --    & 20.02 (0.25) &  19.16 (0.06) &   18.72 (0.12) &   3\\
12/02/07 & 54144.12  & $\geq$ 20.87  & 20.01 (0.13) &  18.98 (0.17) &    --   &    4\\
12/02/07 & 54144.59  &   --    & 20.15 (0.11) &  19.18 (0.03) &   18.72 (0.15) &   2\\
13/02/07 & 54144.91  & $\geq$ 20.60  & 20.21 (0.21) &  18.97 (0.05) &    --   &    4\\
13/02/07 & 54145.03  & $\geq$ 21.61  & 20.19 (0.15) &  19.09 (0.16) &    --   &    4\\
16/02/07 & 54147.92  & $\geq$ 20.65  & 19.97 (0.18) &  19.02 (0.16) &    --   &    4\\
16/02/07 & 54148.05  & $\geq$ 20.09  & 20.11 (0.40) &  19.11 (0.18) &    --   &    4\\
22/02/07 & 54154.05  & $\geq$ 20.57  & 20.08 (0.22) &  19.24 (0.13) &    --   &    4\\
08/03/07 & 54168.07  &   --    & $\geq$ 19.83 &  19.45 (0.13) &     --   &    4\\
11/03/07 & 54171.57  &   --    & 20.20 (0.26) &  19.48 (0.20) &   18.85 (0.02) &    5\\
21/03/07 & 54181.04  & $\geq$ 20.16  & $\geq$ 19.99 &  19.53 (0.35) &     --   &    4\\
24/03/07 & 54183.95  & $\geq$ 20.94  & 20.29 (0.63) &  19.43 (0.03) &     --   &    4\\
15/04/07 & 54206.45  & $\geq$ 21.10  & 20.23 (0.06) &  19.74 (0.14) &   19.26 (0.09) &    1\\ \hline \hline
\end{tabular}
\\
1 = Asiago 1.82m + AFOSC, 2 = LT, 3 = NOT, 4 = FT North, 5 = WHT, 6 = \cite{Nakano06}. \\
}
\end{center}
\end{table*}

\begin{table}
\begin{center}{
\caption[Magnitudes of the sequence stars in the fields of SN 2006ov]{Magnitudes of the stars of the local sequences.  
The numbers in brackets are the r.m.s. of the
available measurements. If no error is reported, only 
a single estimate is available. Due to the large r.m.s. star n.3 was not used for the calibration. } \label{ph_seq_06ov}
\tiny
\begin{tabular}{ccccc}\\ \hline 
Star & B & V & R & I  \\ \hline \hline  
\multicolumn{5}{c}{SN 2006ov} \\ \hline

1 & 15.22 (0.01) & 14.27 (0.01) & 13.65 (0.01) & 13.14 (0.02)  \\ 
2 & 14.56 (0.02) & 13.98 (0.01) & 13.60 (0.01) & 13.24 (0.04) \\
3 & 15.98 (0.58) & 15.43 (0.70) & 14.92 (0.58) & 14.58 (0.56)\\ 
4 & 18.26 (-) & 16.95 (-) & 16.17 (-) & 15.48 (-) \\
5 & 18.78 (0.03) & 17.55 (0.01) & 16.80 (0.02) & 16.07 (0.03) \\
6 & 18.23 (0.04) & 17.59 (0.02) & 17.24 (0.04) & 16.84 (0.02) \\
7 & 19.43 (0.03) & 18.10 (0.04) & 17.07 (0.01) & 16.19 (0.01) \\
8 & 16.68 (0.02) & 15.59 (0.01) & 14.85 (0.01) & 14.19 (0.01) \\ \hline\hline
\end{tabular}
}

\end{center}
\end{table}

\subsection{Colour curves} \label{sec:colour}

In this section we compare the colour curves of an extensive sample of underluminous SNe IIP,
including the objects discussed in the previous sections (SN 2002gd, SN 2003Z, SN 2004eg, SN 2006ov)  together with 
SN 1994N, SN 1999br, SN 1999eu, SN 2001dc (\citealp{Pastorello04}), SN 1997D (\citealp{Turatto98,
Benetti01}), SN 2005cs (\citealp{Pastorello06, Tsvetkov06, Brown07,
Pastorello09}), SN 2008bk (\citealp{VanDyk12}), SN 2008in (\citealp{Roy11}), SN 2009N (\citealp{Takats13}), 
SN 2009md (\citealp{Fraser11}) and SN 2010id (\citealp{GalYam11}).

The explosion epochs for the whole SN sample were computed mainly using the available 
photometric information, and dating the available SN spectra.
In the case of SN 2003Z and SN 2002gd, both well sampled during early stages, 
the explosion epochs have been calculated by comparing the evolution of the light 
curves with those of SN 2005cs. Additional information is provided from the early rising branch
and pre-SN detection limits. In the case of  SN 2004eg and SN 2006ov, for which we 
have a less complete photometric coverage, we dated the explosions making use of  the homogeneity 
of the (B-V) and (V-R) colour  curves  (\citealp{Pastorello04})
and of  the spectroscopic evolution (Sect. \ref{sec:spec}).

\begin{figure}
\begin{center}

\includegraphics[angle=270,width=8cm]{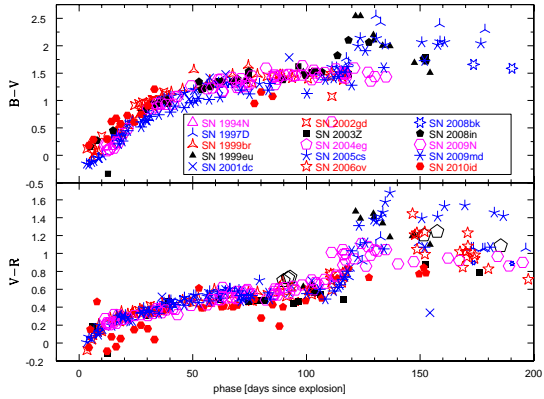}
\caption[bv light curve]{Evolution of (B-V) and (V-R) colours during the photospheric phase for the SNe of our sample.} 
\label{color_lc}

\end{center}
\end{figure}

The (B-V) and (V-R) colour curves during the first 120 days are shown in figure \ref{color_lc}.
Only corrections for galactic extinctions  have been  applied to the whole sample except for the 
cases of SN 2005cs in which the (small) host galaxy extinction is well constrained (A$_{B,tot}$ = 0.205 mag, see \citealp{Pastorello09}) 
and SN 2001dc whose colours indicate significant interstellar reddening. 
This feature, together with the fact that SN 2001dc exploded in a dusty region of the host galaxy, 
lead us to correct its colour for internal extinction as in \cite{Pastorello04} (A$_{B,i}$ = 1.654 mag, A$_{B,tot}$ = 1.693 mag).

During the photospheric phase, following the envelope expansion and cooling,  the (B-V) colour reddens 
during the first 60 days reaching (B-V) $\simeq$ 1.4 mag, then it remains around this value for the subsequent 
 60 days (when the H envelope recombines). The (V-R) colour, reaches $\simeq$ 0.6 mag at about 70 days, followed by a flattening in the subsequent period. A red spike is visible at $\sim$120 days. 

The high degree of homogeneity in the colours of these objects during the first 100 days
might suggest some homogeneity in the properties of the progenitor stars.
 
At about 4 months past core-collapse a larger dispersion in the (B-V) colour is visible among
 faint IIP SNe. A similar behavior is observed for the (V-R) colour, ranging from (V-R) $\simeq$ 0.4 mag in the case
of SN 2004eg to (V-R) $\simeq$ 0.8 mag in the case of SN 2006ov.
This is mostly  due to the intrinsic faintness of the SNe in this phase, that increases significantly the
photometric errors.

\subsection{Bolometric light curves} \label{sec:bolom}

\begin{figure}
\begin{center}
\includegraphics[angle=270,width=8cm]{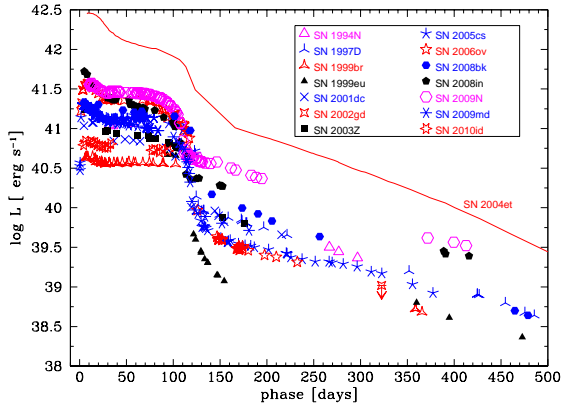}
\caption[BVRI bolometric light curves.]{BVRI bolometric light curves for our LL SN IIP sample.} \label{bol_lc}
\end{center}
\end{figure}

Pseudo-bolometric light curves for our SN sample were obtained by integrating the fluxes in the B, V, R, I bands (Fig. \ref{bol_lc}). 
As a reference object, we also consider the  BVRI pseudo bolometric light curve of the normal type IIP SN 2004et.
It is evident that the bolometric luminosities of all objects in our sample are systematically fainter 
than that of SN 2004et, and - in particular - their plateau luminosities are at least a factor $\times$10 fainter.
As mentioned in previous sections, the plateau phase in sub-luminous SNe IIP lasts about 100-110 days, 
followed by an abrupt drop by about 3-5 mag, when the SN light curves settle onto the much slower radioactive tail powered primarily by the radioactive decay of Co to Fe. 

Despite the general homogeneity in the photometric properties, some
interesting and significant differences are visible among the objects.

The luminosity of the plateau in our sample range between the upper value of SN 2009N 
($L_{BVRI} \approx 2.8 \times 10^{41}$ erg s$^{-1}$), and the lower extreme marked by SN 1999br,  $L_{BVRI} \approx 3.5 \times 10^{40}$ erg s$^{-1}$. 
This wide range in the plateau luminosity suggests some differences in the kinematics of the expanding gas, 
in the initial radius or in the masses of the hydrogen envelope, where the recombination takes place (or a combination of them).

Between days 150 and 500 all light curves show a similar behaviour, with nearly linear luminosity declines, 
consistent with that expected from the radioactive decay of $^{56}$Co into $^{56}$Fe (0.98 mag/100days).
 During this phase, at around 400 days,  we observe a spread in the luminosity of the sample between the lower value of SN 1999br and SN 1999eu ($L_{BVRI} \approx 4 \times 10^{38}$ erg s$^{-1}$) 
and the higher value of SN 2009N ($L_{BVRI} \approx 4 \times 10^{39}$ erg s$^{-1}$). This is an indication of a significant spread in the $^{56}$Ni  masses ejected by LL SNe IIP.

Through a comparison with the late-time luminosity of SN 1987A we could estimate  the $^{56}$Ni  masses for all objects of our SN sample. 
We found values spanning from $1 \times 10^{-3}$ up to $ 2 \times 10^{-2}$ M$_{\odot}$(Ni), 1-2 orders of magnitude less than in normal 
type IIP SNe (for more details, see Sect. \ref{sec:sys}).

\section{Spectroscopy} \label{sec:spec}

In this section we present the spectroscopic evolution of the five under-luminous SNe IIP introduced in the previous sections. 
SN 2002gd, SN 2003Z and SN 2006ov have been well monitored, whilst for SN 1999gn and SN 2004eg only one and two spectra are available, respectively. The journal of spectroscopic observation in presented in Tab. \ref{obs_sp}.

\subsection{Individual properties}
\begin{itemize}

\item{\bf SN 1999gn -} Only one spectrum of SN 1999gn is available (Fig. \ref{seq_spec_04eg_99gn}, top)  taken during the plateau phase. By comparing this spectra with a library of supernova spectra via GELATO (\citealp{Harutyunyan08}) we could estimate a phase of about 45 days past explosion. The continuum is red and is characterized by the presence of a number of P--Cygni lines, including H${\alpha}$,  Sc II 5527 {\r{A}}, Na I D 5890, 5896 {\r{A}}, Ba II (multiplet 1 at 4554, 4934 {\r{A}} and multiplet 2 at 5854, 6142, 6497 {\r{A}}), Fe II,  O I, Ca II, Ti II and Cr II (see \citealp{Pastorello04} for a detailed line identification).
The velocity inferred from Sc II 6246 {\r{A}} line is v (Sc II) = 1690 $\pm$ 100 km s$^{-1}$, significantly lower than that observed in normal type II SNe, and close to the value inferred for SN 2005cs at a similar epoch (v = 1550 km s$^{-1}$; \citealp{Pastorello09}).  

\begin{figure}
\begin{center}
\includegraphics[width=8cm]{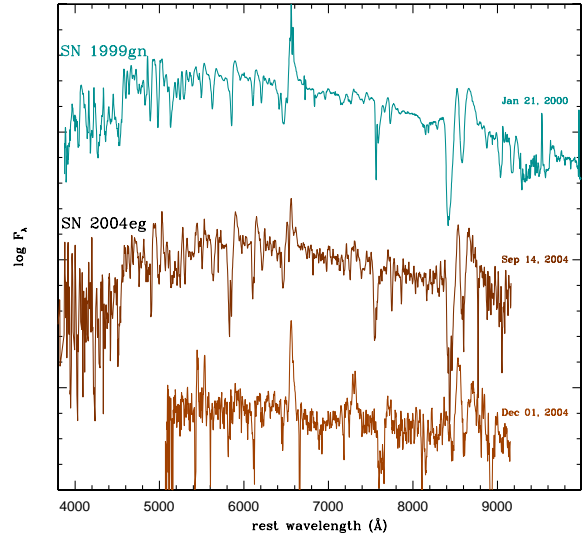}
\caption[Photospheric spectra of SN~1999gn and SN 2004eg]{Spectra of SN 1999gn (top) and SN 2004eg (centre and bottom).} \label{seq_spec_04eg_99gn}
\end{center}
\end{figure}

 \item{\bf SN 2002gd -} We extensively monitored SN 2002gd, especially during the photospheric phase (Fig. \ref{seq_spec_02gd}). The two earliest spectra, obtained with the 
   ESO 4m--class telescopes have the best S/N ratio and are dominated by a blue continuum and only Balmer lines and He I $\lambda$5876 {\r{A}}  
   are detected. In addition, a P--Cygni profile is detected at about 6250 {\r{A}}, possibly identified as Si II 6355 {\r{A}}.   When the temperature fades, Fe II and Ca II lines begin to appear
   and He I is replaced by the Na I doublet. Between days 30 and 40 after explosion,
   other metal lines with P--Cygni profiles become prominent, including O I, Ti II, Sc II, Sr II, Ba II, Cr II.
   The line blanketing at the short wavelengths increases and the continuum becomes redder: this is consistent with the color evolution discussed above. In late--photospheric spectra, H${\alpha}$, Na I D and Ca II are the most prominent spectral features.

\begin{figure}
\begin{center}

\includegraphics[width=8cm]{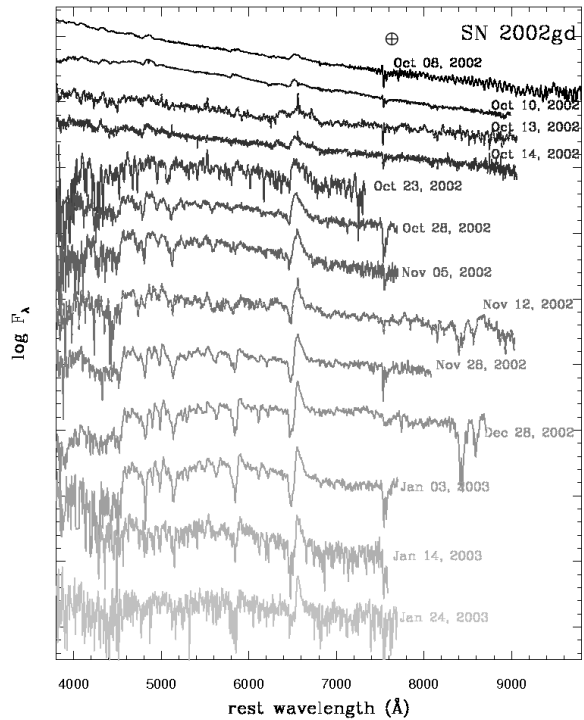}
\caption[Photospheric spectra of SN 2002gd]{Sequence of spectra of SN~2002gd during the photospheric phase. Symbol $\oplus$ indicates telluric absorptions.}  \label{seq_spec_02gd}
\end{center}
\end{figure}

 \item{\bf SN 2003Z -} The first four spectra presented in Fig. \ref{seq_spec_03Z} are from \cite{Knop07} and display the early evolution of this SN. 
We started our monitoring about  $\sim$ 1 month after the discovery.
The early spectra show relatively  strong P--Cygni lines of 
 H, Ca II, Na I, Fe II, while other metal lines,
 such as Sc II, Ba II, Cr II and Ti II become prominent during the plateau phase.
 The sequence of spectra of this SN is one of the most comprehensive
 among LL events. It covers
 the transition from  early spectra (similar to those of canonical
 type II SNe) to  the end of the plateau. At which point the spectra
  have red continua and prominent, very narrow
P--Cygni lines. 
 
\begin{table*}
\begin{center}{
\caption[Journal of spectroscopic observations]{Journal of spectroscopic observations.} \label{obs_sp}
\scriptsize
\begin{tabular}{ccccccc} \\ \hline
Date & JD (2400000+) & Phase (days) & Instrument & Grism & Range (A) & Res. (A) \\ \hline\hline
\multicolumn{6}{c}{SN 1999gn} \\ \hline
21/01/2000 & 51564.77 & 45.7 & NTT+EMMI & gm2 + gm4 & 3900-10600&24+37\\
\hline \hline \\
\multicolumn{6}{c}{SN 2002gd} \\ \hline
08/10/2002 & 52555.53 & 3.5 &  ESO 3.6M+EFOSC2 & gm11 + gm 12 & 3330--9990 & 12+11 \\
10/10/2002 & 52557.69 & 5.7 & ESO NTT+EMMI & gm3 & 3720--9060 & 8 \\
10/10/2002 & 52558.08 & 6.1 & SSO 2.3m+DBS & - &  3300--4270,5790--6750 & 1.1+1.1 \\
13/10/2002 & 52561.06 & 9.1 & SSO 2.3m+DBS & - &  3040--9140 & 5+8 \\
14/10/2002 & 52562.03 & 10.0 & SSO 2.3m+DBS & - &  3160--9140 & 5+8 \\
23/10/2002 & 52571.0  & 19.0 & MSO 74in+B$\&$C & 300 l/mm grating &  3870--7390 & 8 \\
28/10/2002 & 52576.37 & 24.4 & Asiago 1.82m+AFOSC & gm4 &  3400--7760 & 24 \\
05/11/2002 & 52584.33 & 32.3 & Asiago 1.82m+AFOSC & gm4 &  3390--7770 & 24 \\
12/11/2002 & 52590.76 & 38.8 & SSO 2.3m+DBS & - &  3450--9110 & 5 \\
28/11/2002 & 52607.46 & 55.5 & CAHA 2.2m+CAFOS & B200 &  3200--8150 & 13 \\
28/12/2002 & 52637.31 & 85.3 & CAHA 2.2m+CAFOS &  B200 & 3240--8780 & 13 \\
03/01/2003 & 52643.26 & 91.3 & Asiago 1.82m+AFOSC & gm4 &  3390--7760 & 24 \\
14/01/2003 & 52654.27 & 102.3 & Asiago 1.82m+AFOSC & gm4 &  4310--7650 & 24 \\
24/01/2003 & 52664.27 & 112.3 & Asiago 1.82m+AFOSC & gm4 &  3400--7760 & 24 \\ 
\hline \hline \\
\multicolumn{6}{c}{SN 2003Z} \\ \hline
21/02/2003 & 52692.48 & 27.5 & TNG+LRS & LR-B &  3160--8050 & 16 \\
24/02/2003 & 52694.54 & 29.5 & Asiago 1.82m+AFOSC & gm2 + gm4 &  3400--9070 & 24+38 \\
04/03/2003 & 52702.57 & 37.6 & Asiago 1.82m+AFOSC & gm4 &  3580--7820 & 24 \\  
06/03/2003 & 52704.51 & 39.5 & Asiago 1.82m+AFOSC & gm4 &  3590--7810 & 24 \\  
08/04/2003 & 52738.36 & 73.4 & Asiago 1.82m+AFOSC & gm4 &  3560--7760 & 24 \\  
10/04/2003 & 52739.55 & 74.6 & TNG+LRS & LR-B & 3190--8030 &  16 \\
01/05/2003 & 52761.49 & 96.5 & TNG+LRS & LR-B & 3200--8030 &  16 \\
08/05/2003 & 52768.45 & 103.5 & Asiago 1.82m+AFOSC & gr4 &  3410--7750 & 24 \\  
27/06/2003 & 52818.41 & 153.4 & TNG+LRS & LR-B & 3200--8030 &  16 \\
\hline \hline \\
\multicolumn{6}{c}{SN 2004eg} \\ \hline
14/09/2004 & 53263 & 93 & DuPont + WFCCD & - & 3800-9200&-\\
01/12/2004 & 53341.56 & 171.6 & TNG + Dolores & LR-R & 4886-9016&12 \\
\hline \hline \\
\multicolumn{6}{c}{SN 2006ov} \\ \hline
28/11/2006 & 54067.72 & 93.7 & Asiago 1.82m+AFOSC & gm2+gm4 & 3400--9070 & 24 + 38\\
20/12/2006 & 54089.71 & 115.7 & Asiago 1.82m+AFOSC & gm2+gm4 & 3400--9070 & 24 + 38 \\
26/12/2006 & 54095.75 & 121.7 & WHT + ISIS & R158R+R300B & 3076-9597 & 6+9 \\
11/02/2007 & 54142.73 & 168.7 & NOT + ALFOSC & gm5 & 4906-9759 &16\\
12/03/2007 & 54171.72 & 197.7 & WHT + ISIS & R158R+R300B & 3048-10458 & 6+9 \\
\hline \hline \\
\end{tabular}

} 
\end{center}
\end{table*}

\begin{figure}
\begin{center} 
\includegraphics[width=8cm,angle=0]{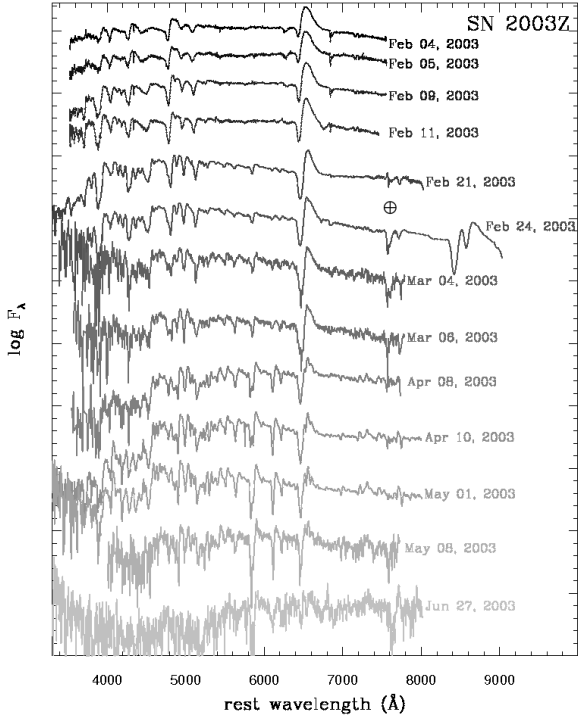}
\caption[Spectral evolution of SN 2003Z]{Spectral evolution of SN 2003Z. Symbol $\oplus$ indicates telluric absorptions.} \label{seq_spec_03Z}
\end{center}
\end{figure}

\item{\bf SN 2004eg - } Two spectra are available for SN 2004eg (Fig. \ref{seq_spec_04eg_99gn}), one at the end of the plateau phase and one in the nebular phase. The first spectrum (+93 days) shows extremely  narrow lines with P--Cygni profiles: Sc II 5527 {\r{A}}, Ba II 6142 {\r{A}}, Na I 5890, 5896 {\r{A}} and Ca II H \& K and near--infrared triplet (8498, 8542, 8662 {\r{A}}) are detected.    
In the nebular spectrum (+171 days) besides H${\alpha}$ and Na I D the most prominent lines are identified as [Ca II] 7291, 7324 {\r{A}}, O I 7774 {\r{A}}  and Ca II near--infrared triplet. H${\alpha}$ and Ca II lines present asymmetric profiles with two components: one with an higher Full Width at Half Maximum (FWHM), from which we inferred a 
velocity v $\sim$ 1400-1500 km s$^{-1}$, and a  narrow marginally resolved component of lower velocity, $v\sim$ 600-850 km s$^{-1}$. 
We note that the [O I] lines (6300, 6364{\r{A}}) are very weak  in the
Dec.1 spectrum. \cite{Jerkstrand12} and \cite{Maguire12} suggest that
high progenitor mass stars ($\sim$ 19 M$_{\odot}$) produce about a
factor two more oxygen during their nucelosynthesis than low mass ones
(M $\leq$ 15 M$_{\odot}$), hence  weakness (or total absence) of these
lines  supports the arguments for a lower mass progenitor
 for SN 2004eg (see sect. \ref{Hydro}).

\item{\bf SN 2006ov -} We collected five spectra of SN 2006ov (see Fig. \ref{seq_spec_06ov}): one at the end of the plateau phase (+93d), two during the post plateau drop (+115d and +121d) and two in the nebular phase (+168d and +197d). 

The continuum at 93 days is red and many narrow P--Cygni lines are visible: H Balmer lines, O I 7774, Ca II, Ba II (multiplet 1 at 4554, 4934 {\r{A}} and multiplet 2 at 5854, 6142, 6497 {\r{A}}), Sc II, Ti II and Cr II (for details see section \ref{com_spec}), along with  Na I D absorption features at 5890, 5896 {\r{A}}. 
The velocities inferred from H$\alpha$ and Sc II lines in this phase
are v(H${\alpha}$) $\sim$ 3320 km s$^{-1}$ and v(Sc II) $\sim$ 1310 km
s$^{-1}$ (Tab. \ref{vel}). The velocity v(H${\alpha}$) is significantly
lower than those measured in normal type IIP SNe, but at the same time
is relatively high when compared with those of other LL SNe IIP at a
similar phase (v(H${\alpha}$) = 1070, 1190, 1130 km s$^{-1}$ for SN
1999eu, SN 2001dc, SN 2003Z, respectively). This effect in SN2006ov 
is probably due to H$\alpha$ being blended with other lines, including Ba II 6497 {\r{A}}. 

The spectra in the post plateau phase show even redder continua, in agreement with the colour evolution discussed above. In this phase the gradual fading of permitted P--Cygni metal lines is observed, with forbidden emission lines, such as [O I], [Fe I], [Fe II], [Ca II], becoming more prominent with time. 

As the supernova ages, the main spectral features become [Ca II] doublet 7291, 7323 {\r{A}}, Ca II IR triplet 8498, 8542, 8662 {\r{A}} and [O I] 6300, 6364 {\r{A}}. Noticeably, O I 7774 remains particularly prominent.

\end{itemize}

\begin{figure}
\begin{center}

\includegraphics[width=8cm]{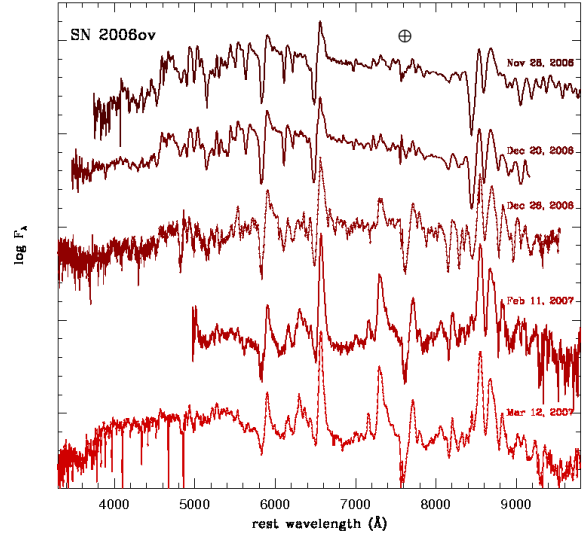}
\caption[Spectra of SN~2006ov]{Sequence of spectra of SN~2006ov from photospheric to nebular phase.Significant contamination from the background explains the blue excess at late phase. Symbol $\oplus$ indicates telluric absorptions.} \label{seq_spec_06ov}
\end{center}
\end{figure}

\subsection{Common spectral properties in LL SNe IIP} \label{com_spec}

On the basis of the data set presented here and in previous papers (\citealp{Turatto98, Benetti01, Pastorello04}, 2006, 2009), we investigate the general spectral properties of an extensive sample of low-luminosity SNe IIP.

\textbf {Phase $\lesssim$ 30d.}
The evolution of low-luminosity type IIP SNe during the early photospheric phase is well illustrated by the spectral sequences of SN 2002gd and SN 2003Z. 
During the first 2-3 weeks, the  spectra of LL SNe IIP are dominated by a blue continuum and only Balmer H lines and He I 5876{\r{A}}~are detectable, showing relatively weak and shallow P--Cygni profiles. Afterwards, Ca II H\&K lines (3934, 3968 {\r{A}}), Fe II (especially the lines of the multiplet 42 at 4924, 5018 and 5169 {\r{A}}), O I 7774 {\r{A}} and the Ca II infrared triplet (8498, 8542, 8662 {\r{A}}) become prominent, while He I 5876{\r{A}} disappears being   replaced by
Na I D (5890, 5896 {\r{A}}). The qualitative spectral evolution of
these LL SNe during this period is similar to that of SNe IIP with
higher luminosities and kinetic energies.

\begin{figure}
\begin{center}
\includegraphics[width=6cm,angle= 270]{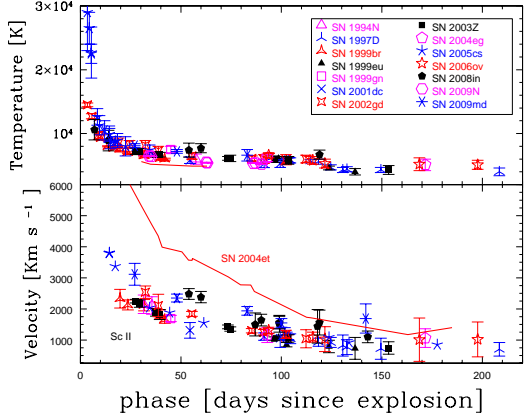}
\caption[Velocities and continuum temperature]{Sc II 6246{\r{A}}~velocities and continuum temperature evolution of LL type IIP SNe.} \label{vel}
\end{center}
\end{figure}

We measured the expansion velocities of the ejecta from the position of the absorption minima of both $H{\alpha}$ and Sc II 6246 {\r{A}} lines, and the temperatures via blackbody
fits to the spectral continuum over regions not affected by line blanketing. The inferred values are reported in Table \ref{vel_T}.
During the first weeks after core-collapse the photospheric velocity,
inferred from the Sc II 6246{\r{A}} minimum (Fig \ref{vel}, bottom
panel), rapidly decreases, reaching $\sim$4000 km s$^{-1}$ at about
two weeks, and $\sim$2000 km s$^{-1}$ at $\sim$30 days. 
Fig. \ref{vel}  illustrates the expansion velocity of the normal SN IIP
SN 2004et (\citealp{Maguire10}) for comparison.
 It is evident that the expansion velocities of LL SNe are lower at any epoch by a factor 2-3 than those of SN 2004et.

The temperature evolution is shown in  Fig. \ref{vel} (top panel) for the whole sample of LL SNe IIP. As a reference we plot the temperature evolution of the normal type IIP event  SN 2004et (\citealp{Maguire10}). 
With age, the continuum, which is initially very blue, becomes rapidly redder (T $\approx$ 10000 K
at phase 10 days, and T $\approx$ 6000-8000 K at phase 30 days). The evolution of all the objects of the sample is remarkably homogenous and similar to those of SN 2004et. 

\textbf{Phase $\sim$ 30-120d.}
With cooling, the absorption lines become deeper and shift to redder wavelengths.
During the plateau phase, between days $\sim$30 and $\sim$120, narrow metal lines appear, viz. Ba II, Sc II, Fe II, Sr II, Cr II, Ti II (with many multiplets visible below 5400 {\r{A}}). With time the absorption line components become narrower and more prominent, and Ba II lines become among the strongest features in the SN spectrum. 

After the fast drop observed during the earliest period, the expansion velocity settles onto a more gentle decline, with v(ScII) slowing from 2000 to 1000 km s$^{-1}$ between day 30 and $\sim$ 120. A similar behaviour is observed also in the temperature evolution (Fig. \ref{vel}), with temperatures remaining almost constant around 5500-6500 K up to  $\sim$120 days. This temperature is close to the recombination temperature for the hydrogen as the photosphere recedes through the envelope. 

\textbf{Phase $\gtrsim$ 120d.}
The end of the recombination phase corresponds to a further decline in the continuum temperature, which settles at T $\sim$ 4000-5000 K as the supernova enters its nebular phase.

The transition to the nebular phase (well visualized by the spectral sequence of SN 2006ov) is characterized  by the progressive fading in the strength of permitted metal lines and by the appearance of  forbidden emissions: [O I] 6300, 6364 {\r{A}}, [Fe II] 7155 {\r{A}}, [Ca II] 7291, 7324 {\r{A}}.
The nebular spectra are very similar to those observed in normal type IIP: the same forbidden lines 
are visible but with narrower profiles.  

The  expansion velocities of the ejecta after 4-5 months, as measured from the FWHM of the strongest lines, are below 1000 km s$^{-1}$, still a factor 2-3 times lower than those of normal SNe IIP, and they remain roughly constant during the subsequent months. 

\begin{table}
\begin{center}{
\caption[Expansion velocities and continuum temperatures]{Expansion velocities as derived from the position of the minimum of H$\alpha$ and the Sc II 6246{\r{A}} line, and continuum temperatures.} \label{vel_T}
\scriptsize
\begin{tabular}{cccc} \\ \hline
Phase & v(H${\alpha}$) km s$^{-1}$ & v(ScII) km s$^{-1}$ & T$_{cont}$ (K) \\ \hline\hline
\multicolumn{4}{c}{SN 1999gn} \\ \hline
45.7 & 2130 (120) & 1690 (100) & 7500 (500) \\ \hline \hline
\multicolumn{4}{c}{SN 2002gd} \\ \hline
3.5 & 5290 (250) & - & 14500 (300) \\
5.7 & 5140 (180) & - & 12700 (400) \\
9.1 & 5010 (330) & - & 9700 (600) \\
10.0 & 4920 (250) & - & 9400 (400) \\
19.0 & 4680 (340) & - & 8500 (800) \\
24.4 & 4480 (170) & - & 8300 (600) \\
32.3 & 4160 (200) & 2540 (200) & 8100 (800) \\
38.8 & 3990 (170) & 2110 (360) & 7300 (700) \\
55.5 & 3580 (170) & 1840 (80) & 6500 (500) \\
85.3 & 2840 (350) & 1300 (70) & 6100 (400) \\
91.3 & 2680 (320) & 1250 (90) & 6000 (700) \\
102.3 & 2530 (450) & 1180 (130) & 5900 (800) \\
112.3 & 2370 (500) & 1040 (290) & 6000 (600) \\ \hline \hline
\multicolumn{4}{c}{SN 2003Z} \\ \hline
27.5 & 4380 (260) & 2250 (60)  & 7200 (250) \\
29.5 & 4260 (230) & 2170 (120) & 7200 (450) \\
37.6 & 3980 (300) & 1890 (120) & 6700 (350) \\
39.5 & 3900 (300) & 1820 (160) & 6700 (350) \\
73.4 & 1660 (100) & 1420 (60) &  6100 (300)\\
74.6 & 1520 (70)  & 1350 (70) & 6100 (500) \\
96.5 & 1130 (60)  & 1040 (60) & 6000 (500) \\
103.5 & 1070 (100) & 1050 (120) & 5800 (600) \\
153.4 & 810 (100) & 740 (200) & 4300 (700) \\ \hline \hline
\multicolumn{4}{c}{SN 2004eg} \\ \hline
93 & 1220 (70) & 1160 (240) & 6200 (600) \\
171.6 & 1440 (410) & 1060 (300) & 5100 (900) \\ \hline \hline
\multicolumn{4}{c}{SN 2006ov} \\ \hline
93.7 & 3320 (470) & 1310 (100) & 6600 (400) \\
115.7 & 2980 (580) & 1090 (160) & 5800 (600) \\
121.7 & 2800 (600) & 1030 (400) & 5300 (700) \\
168.7 & 1920 (430) & 1010 (700) & 5200 (1000) \\
197.7 & 1764 (660) & 1020 (560) & 5100 (700) \\
\hline \hline \\

\end{tabular}

} 
\end{center}
\end{table}

\section{Systematics} \label{sec:sys}

Core collapse SNe show a wide variety of properties depending on the configuration of their progenitor stars at the explosion. 
Even within the subclass of H-rich SNe (IIP)  the observed 
parameters cover a wide range of values.
 \cite{Hamuy03} found that  the physical parameters of SNe IIP, in particular the average plateau 
luminosity, the expansion velocity measured at the recombination, the ejected $^{56}$Ni mass, the total mass of the ejecta and the kinetic energy, are well correlated. 
However, the analysis of \cite{Hamuy03} was biased toward bright objects,
and LL SNe IIP were under-sampled. 
Based on only five objects, \cite{Pastorello04} indicated that LL SNe IIP are not an independent class of core--collapse SNe but rather the low velocity, low $^{56}$Ni mass, low energy tail in the distribution 
of type IIP SNe. Here we want to verify this issue on a much larger sample. 

During the nebular phase, the SN luminosity is powered by radioactive decay chain $^{56}$Ni $\rightarrow$ $^{56}$Co $\rightarrow$ $^{56}$Fe, and its subsequent deposition of $\gamma$-rays and positrons. The ejecta is still opaque to $\gamma$-rays therefore the bolometric luminosity at late phases
can be considered a good indicator of the amount of $^{56}$Ni mass ejected in the explosion. Since typically only optical
band observations are available, the pseudo-bolometric light curve obtained integrating the fluxes over several
passbands (BVRI) can be used as a first approximation to estimate the $^{56}$Ni mass.
Comparing the late-time pseudo-bolometric luminosities of our SN sample with those of SN 1987A in the same 
bands and at the same epochs, we can estimate the  $^{56}$Ni mass ejected by our sample of LL SNe IIP via the relation:

\begin{equation}
M_{SN}(Ni)=0.075 \times \frac{L_{SN}}{L_{87A}} M_{\odot}. 
\end{equation}

A summary of the $^{56}$Ni masses as derived from the above relation is reported in Tab. \ref{Nimass}, along with 
other observables for our SN sample. In the case of SN 2002gd only upper detection limits are available and 
therefore only an upper limit on the ejected  $^{56}$Ni mass is calculated. Moreover, in the case of SN 2005cs  we noticed that the ratio optical/infrared fluxes in the nebular stages is significantly different from that of SN 1987A (this means that considering the BVRI luminosity contribution only does not give a good tool to evaluate the Ni mass of SN 2005cs) . For this reason, in the case of SN 2005cs, we derived the nickel mass by comparing the quasi-bolometric (UVOIR) luminosities instead of the BVRI luminosities. 

\begin{figure}
\begin{center}

\includegraphics[width=6cm,angle= 270]{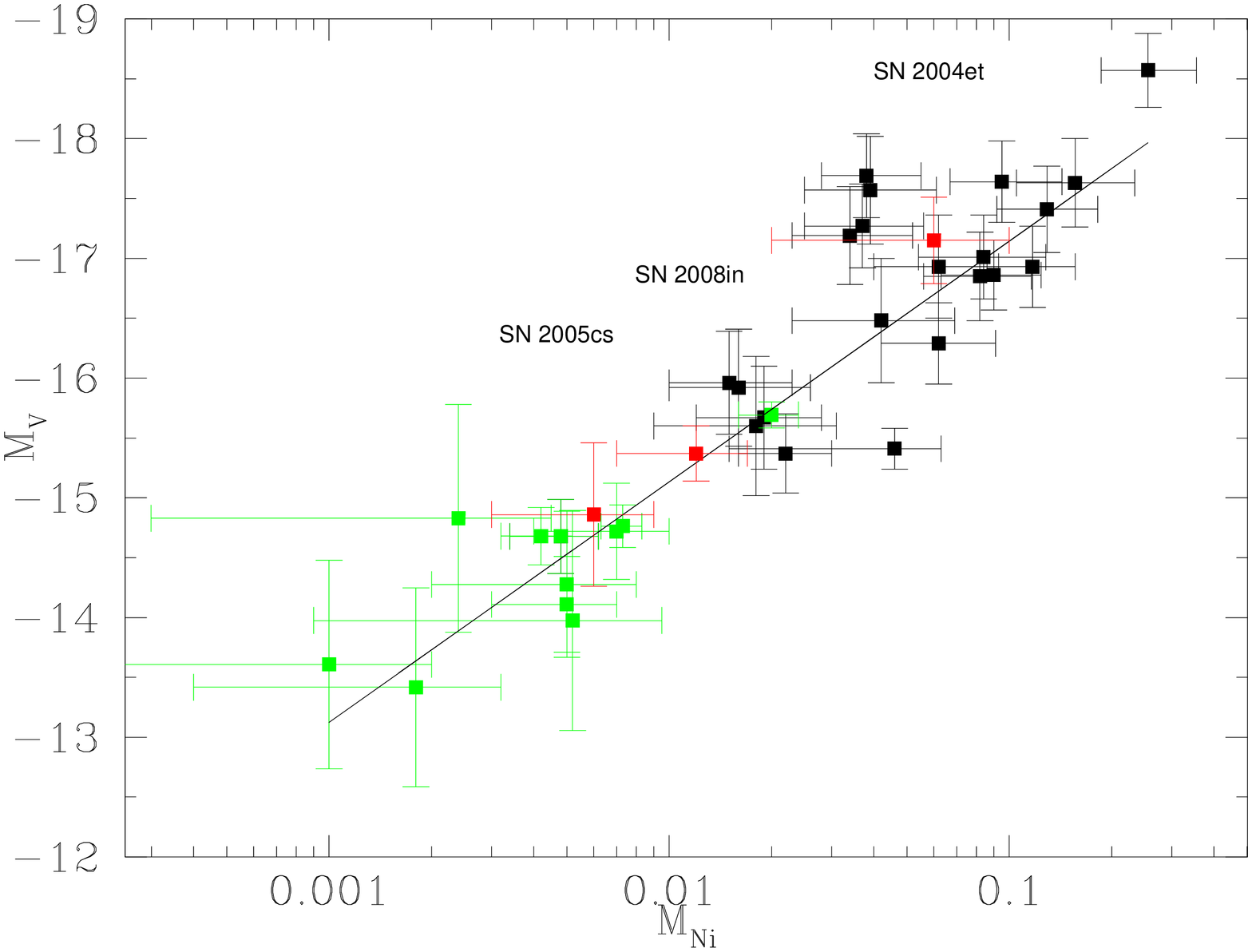}
\caption[$^{56}$Ni masses vs plateau magnitudes V ]{Absolute V-band magnitude (computed at day 50) vs mass of $^{56}$Ni. Green squares represent our LL SN IIP sample, black squares are canonical SNe IIP from \cite{Hamuy03}, \cite{Hendry05,Hendry06}, \cite{Maguire10}, red squares are SN 2004et, SN 2005cs, SN 2008in (\citealp{Roy11}). When no other distance determinations are available, we computed a kinematic distance using the host galaxy radial velocity corrected for the Local Group infall into Virgo (V$_{vir}$ as reported in the HYPERLEDA database, with Hubble constant H$_0$=72 km s$^{-1}$ Mpc $^{-1}$). The adopted uncertainty in the kinematic distances is computed from the local cosmic thermal velocity 187 km s$^{-1}$, as in \cite{Smartt09}. For SN 2004et, we adopted $t_0$ = 2453270, A$_v$ = 1.271, $\mu$ = 28.85 $\pm$ 0.34 (see Tab. 13).} 
\label{Mni_MV}
\end{center}
\end{figure}

We investigate the correlation between the absolute magnitude of the plateau in the V band and the ejected $^{56}$Ni mass derived by Hamuy (2003, see Fig. \ref{Mni_MV}), considering the entire sample of LL SNe IIP (green squares) and 
an extensive sample of normal to luminous SNe IIP (black squares), the SN sample presented by \cite{Hamuy03}, plus the normal type IIP SNe 2003gd (\citealp{Hendry05}), 2004A (\citealp{Hendry06}) and 2004et (\citealp{Maguire10}). We plot with red squares SN 2004et as representative of a normal SNe IIP, SN 2005cs as representative of LL SNe IIP and SN 2008in (\citealp{Roy11}) as an intermediate case (see  section \ref{Hydro}). 
LL SNe IIP appear to fill the observational gap between the ultra-faint SN 1999br (\citealp{Pastorello04}) and the normal type IIP events. 
In spite of a large dispersion, LL SNe IIP do not lie in a separate
area of the diagram and, confirming the result of
\cite{Hamuy03}, we observe  a continuum distribution in the ejected $^{56}$Ni mass with the SN luminosity in our enlarged sample. 
Using a weighted least-squares method, we estimate a correlation coefficient  of  $-0.72$ and $-0.92$ for the normal and entire sample respectively. 
 In other words, there is clear evidence that SNe with brighter plateaus produce more $^{56}$Ni.

\begin{figure}
\begin{center}

\includegraphics[width=6cm,angle= 270]{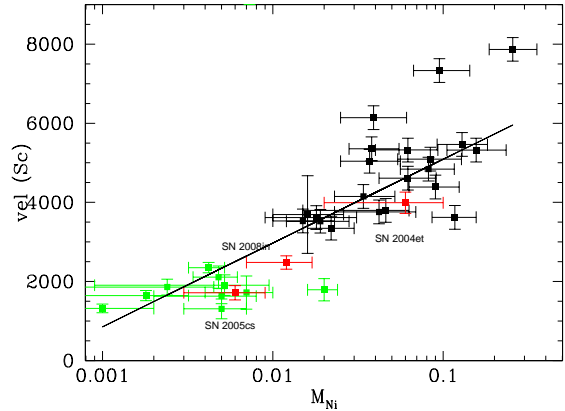}
\caption[$^{56}$Ni masses vs envelope velocities ]{Ejecta velocities vs $^{56}$Ni masses. The expansion velocities are obtained from measuring the position of the minimum of the Sc II 6246 {\r{A}} line. Symbols are as in Fig. \ref{Mni_MV}.} \label{Mni_v}
\end{center}
\end{figure}

In Fig. \ref{Mni_v} we tested the correlation between the photospheric expansion velocities, measured from the position of the minimum of the Sc II $\lambda$6246 line at $\sim$50 days after the explosion, and the ejected $^{56}$Ni mass, extending the original sample of Hamuy to the new faint objects. Again, LL SNe IIP are labeled with green colours and normal-to-luminous SNe IIP with black colours. It is clear that SNe with larger expansion velocities are those producing more $^{56}$Ni, confirming once again the results of \cite{Hamuy03}. Using a weighted least-squares method, we estimate a correlation coefficient of $0.65$ and $0.86$ for the normal and entire sample respectively.

From Figs. \ref{Mni_MV} and \ref{Mni_v},  we confirm that there exists a continuum in the properties of SNe type II from low luminosity, low velocity and Ni- poor events such as SN 1997D and SN 1999br, to high luminosity and high velocity SNe IIP (such as SN 1992am, \citealp{Hamuy03}).

\begin{table*}
\begin{center}{
\caption[Properties of LL SNe IIP ]{Properties of the whole sample of low luminosity type IIP SNe. M$_V$ is the mean magnitude in the plateau phase, M$_{Ni}$ is the ejected nickel mass and  $v_{50} $ is the velocity at $\sim$50 days after the explosion. Properties of the normal IIP SN 2004et are reported for comparison. 
References: (1) \cite{Pastorello04}; (2) \cite{Turatto98}; (3) \cite{Benetti01}; (4) this paper; (5) \cite{Pastorello06}; (6) \cite{Pastorello09}; (7) \cite{Mattila08}; (8) \cite{VanDyk12}; (9) \cite{Roy11}; (10) \cite{Takats13}; (11) \cite{Fraser11}; (12) \cite{GalYam11}; (13) \cite{Maguire10}.}\label{Nimass}
\begin{tabular}{cccccccc} \\ \hline
SN & t$_0$ (240 0000+) & $\mu$ & A$_{v}$ & M$_V$ & M$_{Ni}$& $v_{50} $ & Ref.\\
1994N & 49451$\pm$10 & 33.09 $\pm$ 0.31 & 0.108 & -14.68 $\pm$ 0.31 & 0.005 $\pm$ 0.001&2110 $\pm$ 280&1, 4\\ 
1997D & 50361$\pm$15 & 30.74 $\pm$ 0.92 & 0.058 & -13.97 $\pm$ 0.92 &0.005 $\pm$ 0.004&1910 $\pm$ 260&2, 3, 4\\ 
1999br& 51278$\pm$ 3& 30.97 $\pm$ 0.83 & 0.065 & -13.42 $\pm$ 0.83 & 0.002 $\pm$ 0.001&1640 $\pm$ 130&1\\
1999eu& 51394$\pm$15 & 30.85 $\pm$ 0.87 & 0.073 & -13.61 $\pm$ 0.87 & 0.001 $\pm$ 0.001&1320 $\pm$ 110&1, 4\\
2001dc& 51047$\pm$5 & 32.64 $\pm$ 0.38 & 1.250 & -14.11 $\pm$ 0.40 & 0.005 $\pm$ 0.002&1310 $\pm$ 250&1\\
2002gd& 52552$\pm$2 & 32.87 $\pm$ 0.35 & 0.184 & -15.49 $\pm$ 0.35 & $<$0.003&1840 $\pm$ 80&4\\
2003Z& 52665$\pm$4 & 31.70 $\pm$ 0.60 & 0.106 & -14.28 $\pm$ 0.61 & 0.005 $\pm$ 0.003&1630 $\pm$ 190&4\\
2004eg& 53170$\pm$30 & 32.64 $\pm$ 0.38 & 1.237 & -14.72 $\pm$ 0.40 &0.007 $\pm$ 0.003&1730 $\pm$ 100&4\\
2005cs& 53549$\pm$1 & 29.46 $\pm$ 0.60 & 0.155 & -14.86 $\pm$ 0.60 & 0.006 $\pm$ 0.003&1715 $\pm$ 180&5, 6\\
2006ov& 53974$\pm$6 & 30.5 $\pm$ 0.95 & 0.061 & -14.83 $\pm$ 0.95 &0.002 $\pm$ 0.002&1860 $\pm$ 200&4\\
2008bk& 54550$\pm$1 & 27.68 $\pm$ 0.13 & 0.065 & -14.80 $\pm$ 0.13 & 0.007 $\pm$ 0.001&-& 7, 8\\
2008in& 54825$\pm$1 & 30.60 $\pm$ 0.20 & 0.305 & -15.37 $\pm$ 0.23 & 0.012 $\pm$ 0.005&2480 $\pm$ 170&9\\
2009N& 54848$\pm$1 & 31.67 $\pm$ 0.11 & 0.350 & -15.59 $\pm$ 0.12 & 0.020 $\pm$ 0.004&1790 $\pm$ 280 &10\\
2009md& 55162$\pm$8 & 31.64 $\pm$ 0.21 & 0.310 & -14.68 $\pm$ 0.24 & 0.004 $\pm$ 0.001&2350$\pm$ 130&11\\
2010id& 55452$\pm$2 & 32.86 $\pm$ 0.50 & 0.167 & -13.99 $\pm$ 0.51& - &2040 $\pm$ 180&12\\
\hline \hline \\
2004et& 53270$\pm$1 & 28.85 $\pm$ 0.34 & 1.271 & -17.15 $\pm$ 0.27& 0.056 $\pm$ 0.040&3990 $\pm$ 270&13, 4\\
\hline \hline \\

\end{tabular}

} 

\end{center}
\end{table*}

\section{Discussion}\label{Hydro}

Theory predicts that stars with ZAMS masses between $\sim$8-9 M$_{\odot}$ 
and $\sim$25-30 M$_{\odot}$ end their lives as type IIP SNe (e.g. Heger et al. 2003).
The lower limit for this main sequence mass range is set by the heaviest stars that are expected to produce a white dwarf whilst 
the upper limit depends on the details of the pre-SN evolution and the explosion mechanism, and is deeply related to still uncertain parameters such as overshooting, mixing and mass loss rate (see e.g. \citealp{Woosley02}).

Three different possible configurations for the progenitors have been proposed  to explain the energetics and the overall properties of LL Type IIP SNe: {\bf i)} a red supergiant star of about $\sim$12 M$_{\odot}$ which terminates its life when it exhausts the nuclear fuel and its iron core is  no longer sustained against gravitational collapse; {\bf ii)} an explosion of a more massive (M $\gtrsim25$ M$_{\odot}$) star in which  a considerable fraction of the material ejected in the explosion falls back into the compact remnant (this event is sometimes labelled as {\it fall-back SN}); {\bf iii)} stars in the mass range $\sim$8-11 M$_{\odot}$ that forms a degenerate NeO core during the final stage of its evolution and explodes as an electron capture SN. 

In the last decade major efforts have been devoted to constrain the mass of the progenitors of LL type IIP SNe. Those were based on two different approaches, i.e. by comparing observations (light curves, colours, spectra) with hydrodynamical models and by direct detection of the progenitor stars in pre-SN images. With the latter approach,  after measuring the brightness and colour of the star, the mass can be computed through
a comparison with  stellar evolution models (see \citealp{Smartt09b} for a review).
The mass estimates through the two different approaches have been shown to provide somewhat discrepant results with hydrodynamical modelling providing often higher masses for the precursor of type IIP SNe. 

In this section we aim to derive the  physical parameters of two
representative objects by hydrodynamical modelling of the SN data:  SN
2005cs (which has excellent observational coverage and one can
consider it a template for LL SNe IIP) and SN 2008in (an intermediate object between normal and faint SNe IIP).  
This work is the starting point for a detailed analysis through the modelling of the physical parameters of our full sample of LL SNe II-P that will be presented in a forthcoming paper (Pumo et al. in preparation).

We estimate the main physical properties of the progenitors 
at the explosion (i.e. the ejected mass, the progenitor radius and the explosion energy) 
through the hydrodynamical modelling of their main observables (i.e. bolometric light 
curve, evolution of line velocities and continuum temperature at the photosphere), using 
the same well-tested approach used for other observed CC-SNe (e.g.~SNe 2007od, 2009bw, 
2009E and 2012A; see \citealp{Inserra11,Inserra12}, \citealp{Pastorello12}, and \citealp{Tomasella13}). 

According to this approach, a simultaneous $\chi^{2}$ fit of the aforementioned observables 
against model calculations is performed with two codes: a semi-analytic code (\citealp{Zampieri03}) 
which solves the energy balance equation for a homologously expanding envelope of constant 
density and the general-relativistic, radiation-hydrodynamics Lagrangian 
code presented in \cite{Pumo10} and \cite{Pumo11}. The latter is able to 
simulate the evolution of the physical properties of the CC-SN ejecta and the behaviour of the 
main observables from the breakout of the shock wave at the stellar surface up to the nebular 
stage. The distinctive features of this new code are: {\bf a)} an accurate treatment 
of radiative transfer coupled with relativistic hydrodynamics, {\bf b)} a fully implicit Lagrangian 
approach to the solution of the coupled non-linear finite difference system of relativistic 
radiation-hydro equations, and {\bf c)} a description of the evolution of ejected material which 
takes into account both the gravitational effects of the compact remnant and the heating effects 
linked to the decays of the radioactive isotopes synthesized during the CC-SN explosion.

The semi-analytic code is used to carry out a preparatory study aimed at determining the 
parameter space describing the CC-SN progenitor at the explosion and, consequently, to guide 
the more realistic, but time consuming simulations performed with the general-relativistic, 
radiation-hydrodynamics code.

The approach of modelling with the two codes is appropriate, since the emission of SNe 2005cs 
and 2008in is dominated by the expanding ejecta.
In 
performing the $\chi^{2}$ fit, we have neglected the observations obtained before the full relaxation of the plateau ($t <$ 20d) when 
 the observables are significantly affected by emission from the outermost shell of the ejecta (see \citealp{Pumo11} for details). 
After shock passage, this shell, that contains only a small fraction of the envelope mass (few tenths of solar masses), is accelerated to very high velocities and is not in homologous expansion. The structure, evolution and emission properties of this shell are not well reproduced in our simulations because at present we adopt an "ad hoc" initial density profile, not derived from an explosion simulation coupled with pre-SN evolutionary models.

\begin{figure}
\begin{center}
\includegraphics[width=6cm,angle= 270]{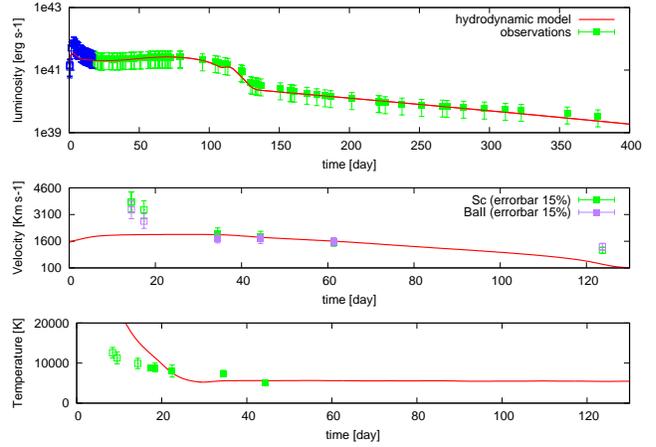}
\caption[modelling of SN 2005cs]{Comparison of the evolution of the main observables of SN 2005cs with the best-fit model                                                            
computed the general-relativistic, 
radiation-hydrodynamics code (total energy $0.16$ foe, initial radius $2.5 \times 10^{13}$ cm, 
envelope mass $9.5$M$_{\odot}$). Top, middle, and bottom panels show the bolometric light curve, the 
photospheric velocity, and the photospheric temperature as a function of time. Blue symbols in the top panel and open symbols in the middle and bottom panels mark early observations ($<$ 20 days) not considered in the fit (see text for details). Photospheric velocities are those derived from the minima of the P--Cygni profiles of the 
Sc II and Ba II lines.} \label{SN05csmod}
\end{center}
\end{figure}

\begin{figure}
\begin{center}
\includegraphics[width=6cm,angle= 270]{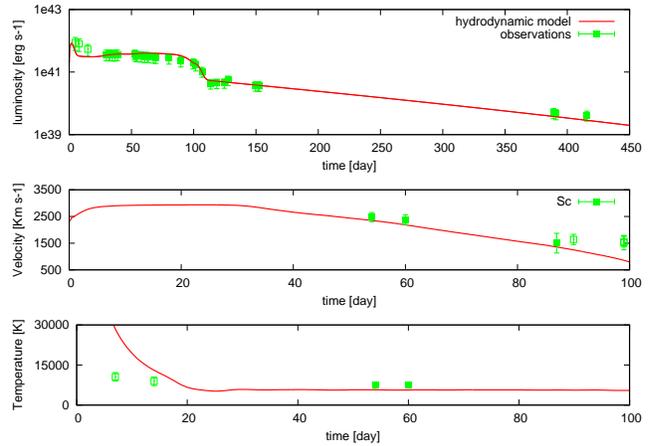}
\caption[modelling of SN 2008in]{Comparison of the evolution of the main observables of SN 2008in with the best-fit model                                                            
computed with the general-relativistic, 
radiation-hydrodynamics code (total energy $0.49$ foe, initial radius $1.5 \times 10^{13}$ cm, 
envelope mass $13$M$_{\odot}$). The three panels are similar to those of Fig. \ref{SN05csmod}.} \label{SN08inmod}
\end{center}
\end{figure}

The best-fitting models for SNe 2005cs and 2008in are shown in Figs.~\ref{SN05csmod} and
\ref{SN08inmod}, respectively.

For SN 2005cs, assuming a $^{56}$Ni mass of $\sim$ 0.0065 M$_{\odot}$, an explosion epoch JD$= 2453549$  and a distance modulus of $29.46\pm0.60$ mag (see Tab.\ref{Nimass}),
the best fit returns values of total (kinetic plus thermal) energy of $E = 0.16$ foe, 
initial radius of $R = 2.5 \times 10^{13}$ cm, and envelope mass of M$_{env}$ = 9.5 M$_{\odot}$, in good agreement with the parameters obtained from our previous modeling reported in \cite{Pastorello09}. The estimated uncertainty on the best-fit model
parameters $E, M_{env}$ and $R$ is $\lesssim$ 20\%. 
The values reported above are consistent with the explosion of a moderate mass star. Indeed, 
adding the mass of the compact remnant ($\sim 1.5$ M$_{\odot}$) to that of the ejected material, we 
obtain that the mass of the progenitor of SN 2005cs at the explosion was $\sim 11$ M$_{\odot}$.
Since the observables of all other LL SNe of our sample are rather similar to those of SN
2005cs, it is reasonable to assume that they share similar physical parameters
as those inferred for SN 2005cs (a detailed analysis will be
reported in Pumo et al., in prep.).

The situation is different for SN 2008in (\citealp{Roy11}) whose observed parameters are intermediate 
between those of canonical SNe IIP and LL SNe IIP. 
As for SN 2005cs, adopting a $^{56}$Ni mass of $\sim$ 0.012 M$_{\odot}$, 
an explosion epoch  JD$= 2454825.6$  and a distance modulus of $30.60\pm0.20$ mag (see Tab.~\ref{Nimass}),
 the best fit  model returns values of total (kinetic plus thermal) energy of $E = 0.49$ foe, 
initial radius of $R = 1.5 \times 10^{13}$ cm, and envelope mass of M$_{env}$ = 13 M$_{\odot}$.
Again, the estimated uncertainty on the best-fit model
parameters $E, M_{env}$ and $R$ is $\lesssim$ 20\%. 
These values are consistent with the explosion of an intermediate mass star, slightly more massive than the progenitor of SN 2005cs.
Adding the mass of the compact remnant to that of the ejected material, we obtain that the mass of 
the progenitor of SN 2008in at the explosion was  of $\sim 14.5$M$_{\odot}$.

Up to now, a relatively small number of LL SNe IIP have been discovered. This is plausibly due to selection effects due to their faint intrinsic luminosity rather than intrinsic rarity.
Among this restricted sample, hydrodynamical models were compared with observables  of five objects  (SN 1997D, SN 1999br, SN 2003Z, SN 2005cs, SN 2008in). 

\cite{Zampieri03} found a relatively massive progenitor (19M$_{\odot}$) for SN 1997D (revised downwards to 14 $\pm$ 2 M$_{\odot}$ in \citealp{Zampieri07}). The low amount of $^{56}$Ni observed  was explained as due to fallback of material onto the collapsed remnant.
 An intermediate mass of 16M$_{\odot}$  for the progenitor was found by \cite{Zampieri03} for SN 1999br (revised to 10 $\pm$ 1.5 M$_{\odot}$ in \citealp{Zampieri07})  and of 15.9 $\pm$ 1.5 M$_{\odot}$ for SN 2003Z by \cite{Utrobin07}.
In the case of SN 2005cs, the best observed LL IIP SN, several groups reported hydrodynamic masses of 18.2 $\pm$ 1 M$_{\odot}$ (\citealp{Utrobin08}), 10-15M$_{\odot}$ (\citealp{Pastorello09}) to be compared with the new value of $\sim$11 M$_{\odot}$ obtained here with the new hydrodynamical model. 
Finally, we find the mass of the progenitor of SN 2008in at the
explosion to be $\sim$ 14.3 M$_{\odot}$, in good agreement with
\cite{Utrobin13} who found a progenitor mass of 15.5 $\pm$ 2.2
M$_{\odot}$ (including the mass loss effects)  for the same object.
The differences among the results obtained with different
hydrodynamical approaches, are likely a consequence of different
assumptions about the input physics (in particular the opacity) and of
different choices of distance moduli or total extintion (e.g. $\mu$ =
29.62 vs $\mu$ = 29.26 and E(B-V) = 0.12 mag vs E(B-V) = 0.05 mag
assumed by \cite{Utrobin08} and \cite{Pastorello09} respectively for
SN 2005cs). Despite these differences, recent results from
hydrodynamical modeling now converge to low--intermediate masses for LL SNe IIP.

Hydrodynamical models have been applied also to well studied normal SNe IIP, e.g. SN 1999em and SN 2004et. 
In the case of SN 1999em a pre-supernova radius of $R \sim (3.5-6.9) \times 10^{13}$ cm, an ejecta mass of 18-19 M$_{\odot}$, an explosion energy of 1 - 1.3 foe and a radioactive $^{56}$Ni mass of 0.036 - 0.06  M$_{\odot}$ were estimated (\citealp{Utrobin07b}, \citealp{Baklanov05}, \citealp{Bersten11}). 
Larger values for the physical parameters were obtained in the case of SN 2004et: $R \sim 10^{14}$ cm, M$_{env}$ = 24.5 M$_{\odot}$, $E \sim 2.3$ foe, M$_{Ni} \sim 0.068$ M$_{\odot}$ (\citealp{Utrobin09}). 

Despite the fact that the values reported above were derived by different hydrodynamical codes with different assumption, physics, etc., we speculate that there could be a general trend in the parameters of type IIP SNe, as claimed by \cite{Hamuy03}.  Less energetic explosions preferentially produce less luminous
events and lower mass of $^{56}$Ni.
Moreover, the parameters inferred from data modeling of LL SNe IIP appear now to agree with the analysis of pre-SN images in suggesting that stars of relatively low-to-intermediate mass produce LL SNe while normal SNe IIP derive from progenitors of larger mass. A more sophisticated hydrodynamical modeling of LL SNe IIP is in progress (Pumo et al. in prep.) to strengthen this speculation.

\section{Summary }

In this paper we have presented new data for five LL SNe IIP and investigated their observational properties, comparing them with a comprehensive sample of objects.
 We have shown that all LL SNe IIP have very homogeneous photometric and spectroscopic evolutions.
In particular they display surprisingly consistent colour evolutions, expansion velocities of the ejected material and amounts of $^{56}$Ni synthesized (see table \ref{Nimass}). We have investigated the correlations of the ejected $^{56}$Ni masses with the plateau magnitudes and expansion velocities of the ejecta, and showed  that these faint events are the low-luminosity tail of a continuous distribution of otherwise normal explosions, instead of a separate class of CC SNe.
Finally we have discussed the preliminary results of a systematic hydrodynamical modelling of the observables of LL SNe IIP that will be presented more extensively in a forthcoming paper. The hydrodynamical masses estimates for SN 2005cs and for SN 2008in suggest that LL SNe IIP originate from low-to-intermediate mass stars in the range 10-15 M$_{\odot}$, in good agreement with the mass trend suggested by the direct progenitor detection method (\citealp{Smartt09}). 

It seems that results from our modeling and direct detection of the progenitor in pre-explosion images  are converging toward a  low-to-intermediate mass scenario (see also \citealp{Tomasella13}). 
Indeed the results presented in this paper for SN 2005cs and SN 2008in reduce the gap between progenitor masses estimated through hydrodynamical modeling or direct detection in pre-SN archive images, pointing to red supergiants of moderate mass or, less commonly, to super AGB stars (although the possibility to have electron-capture SN from super-AGB progenitor is strongly questioned, e.g. \citealp{Eldridge07}). Differences in the light curves or in the spectral properties among the events reported here could be trivially explained with slightly different values of energies, radii or ejected masses.

\section*{Acknowledgements}
Based on observations made with:\\
- ESO Telescopes at the La Silla Paranal Observatory under programme ID 60.A-9013(A), ID 70.B-0338(A), ID 64.H-0467(B); \\
- The Cima Ekar 1.82m telescope of the INAF-Astronomical Observatory of Padua, Italy;\\
- The Liverpool Telescope operated on the island of La Palma by Liverpool John Moores University in the Spanish Observatorio del Roque de los Muchachos of the Instituto de Astrofisica de Canarias with financial support from the UK Science and Technology Facilities Council. \\
-  the Nordic Optical Telescope, operated by the Nordic Optical Telescope Scientific Association at the Observatorio del Roque de los Muchachos, La Palma, Spain, of the Instituto de Astrofisica de Canarias.\\
- The William Herschel and the Jacobus Kapteyn telescopes operated on the island of La Palma by the Isaac Newton Group in the Spanish Observatorio del Roque de los Muchachos of the Instituto de Astrof'sica de Canarias. \\
- The Italian Telescopio Nazionale Galileo (TNG) operated on the island of La Palma by the Fundaci—n Galileo Galilei of the INAF (Istituto Nazionale di Astrofisica) at the Spanish Observatorio del Roque de los Muchachos of the Instituto de Astrofisica de Canarias.

We thank  M.Riello, S.Gagliardi for observations of SN 2002gd and L. Di Fabrizio for observation of SN 2006ov. \\
We also thank R. Roy, K. Maguire and M. Fraser for informations on SN 2008in, SN 2004et  and SN 2009md respectively. \\
We acknowledge the TriGrid VL project and the INAF-Astronomical Observatory of Padua for 
the use of computer facilities. S.S. acknowledges the support of ASI contract n. I/023/12/0.
M.L.P. acknowledges the financial support from the PRIN-INAF 2011 Transient Universe: from ESO Large to PESSTO (P.I. S. Benetti). A.P., M.T. S.B., E.C. and A.H. are also partially supported by the same PRIN.
The research leading to these results has received funding from the European Research Council under the European Union's Seventh Framework Programme (FP7/2007-2013)/ERC Grant agreement n$^{\rm o}$ [291222]  (PI : S. J. Smartt).  
N.E.R. is supported by the MICINN grant AYA2011-24704/ESP, by the ESF EUROCORES Program EuroGENESIS (MICINN grant EUI2009-04170), by SGR grants of the Generalitat de Catalunya, and by EU-FEDER funds.
G.P. received partial support from Center of Excellence in Astrophysics and Associated Technologies (PFB 06)

This research has made use of the NASA/IPAC Extragalactic Database (NED) which is operated by the Jet Propulsion Laboratory, California Institute of Technology, under contract with the National Aeronautics and Space Administration. 
 We acknowledge the usage of the HyperLeda database (http://leda.univ-lyon1.fr).\\

\end{document}